\documentclass[pra,twocolumn,preprintnumbers,floatfix]{revtex4}
\usepackage{graphicx}% Include figure files
\usepackage{dcolumn}% Align table columns on decimal point
\usepackage{multibib}
\newcites{supp}{Supplementary References} % Defines \citesupp %and \bibliographysupp
\usepackage{bm}% bold math
\usepackage{latexsym,epsfig}
\usepackage{graphicx}
\usepackage{verbatim}
\usepackage{comment}
\usepackage{amsmath}
\usepackage{amssymb}
\usepackage{stmaryrd}
\usepackage{color}
\usepackage{epstopdf}
\usepackage{grffile}
\usepackage{ulem}
\usepackage{float}
\DeclareGraphicsExtensions{.eps}

\newcommand{\beq}{\begin{equation}}
\newcommand{\eeq}{\end{equation}}
\newcommand{\bea}{\begin{eqnarray}}
\newcommand{\eea}{\end{eqnarray}}
\newcommand{\ben}{\begin{eqnarray*}}
\newcommand{\een}{\end{eqnarray*}}
\newcommand{\bfig}{\begin{figure}}
\newcommand{\efig}{\end{figure}}

\begin{document}
\title{Quantum Simulation of Detuning-Controlled Proximity-Induced Localization and Re-entrant Delocalization in a Quasiperiodic Ladder}
\author{Mrinal Kanti Giri$^{1}$}
\email{mrinalphy333@gmail.com}
\author{Shih Tai-Ming$^2$}
\author{Pinaki Sengupta$^1$}
\author{Pochung Chen$^{2,3,4}$}
\email{pcchen@phys.nthu.edu.tw}

\affiliation{$^1$School of Physical and Mathematical Sciences, Nanyang Technological University, Singapore 637371}
\affiliation{$^2$Department of Physics, National Tsing Hua University, Hsinchu 30013, Taiwan}
\affiliation{$^3$Frontier Center for Theory, Computation, and Data Science, National Tsing Hua University, Hsinchu 30013, Taiwan}
\affiliation{$^4$National Center for Excellence in Quantum Information Science and Engineering, National Tsing Hua University, Hsinchu 30013, Taiwan}

%\thanks{$*$ mrinalphy333@gmail.com}
%\thanks{$\dagger$ pcchen@phys.nthu.edu.tw }

%\date{\today}
\begin{abstract}
Proximity effects, in which one subsystem imprints its properties onto another, offer a route to controlling quantum matter. For localization, this raises the question of whether an extended system can be localized by its neighbor, and then delocalized again, in the regime of weak quasiperiodic disorder. Here, we demonstrate this behavior in a coupled Aubry--Andr\'e ladder, where a staggered detuning applied to only one leg localizes the neighboring, undetuned leg through inter-leg hybridization, even before the detuned leg itself localizes. Upon further increasing the detuning, the undetuned leg undergoes re-entrant delocalization, while the detuned leg remains localized. Finite-size scaling supports the persistence of these regimes in the thermodynamic limit. On an IBM Quantum processor, we experimentally observe their dynamical signatures: strongly suppressed wave-packet spreading in the proximity-localized regime and its recovery upon re-entrant delocalization. Our results establish leg-selective detuning as a reversible control knob for localization and transport in coupled quantum systems.
\end{abstract}
%\begin{abstract}
%We demonstrate proximity-induced localization and re-entrant delocalization in a detuned quasiperiodic ladder. In the weak-quasiperiodic regime, where uncoupled Aubry--André chains are extended, a staggered detuning applied to only one leg induces localization in the neighboring leg through inter-leg hybridization. With further increasing detuning, the pure chain re-enters a delocalized regime, producing a delocalized--localized--delocalized sequence without directly modifying its quasiperiodic potential. We identify these regimes using static localization diagnostics and demonstrate experimentally accessible dynamical signatures on IBM Quantum hardware: strongly suppressed wave function spreading in the proximity-localized regime and its re-entrant recovery at larger detuning. These results establish leg-selective detuning as a programmable control knob for inducing and reversing localization in coupled quasiperiodic systems.
%\end{abstract}

\maketitle

{\bf{\textit{Introduction.-~}}} Localization in low-dimensional quantum systems has attracted sustained attention since Anderson's seminal work~\cite{Anderson1958}. Arbitrarily weak uncorrelated disorder localizes all eigenstates in one and two dimensions, whereas a genuine localization transition requires higher dimensions~\cite{Abrahams1979,LeeRamakrishnan1985,RevModPhys.80.1355}. Quasiperiodic systems, in contrast, provide a route to a localization transition in one dimension~\cite{Soukoulis1982,PhysRevLett.61.2144,PhysRevA.90.061602,PhysRevA.75.063404,Biddle2010,Ganeshan2015}. In the paradigmatic Aubry--Andr\'e (AA) model, the potential is deterministic, generated by an incommensurate modulation~\cite{AubryAndre1980}, and exact self-duality yields a sharp, energy-independent transition at a critical modulation strength~\cite{Aulbach_2004}. Generalized AA models, with multichromatic modulations or long-range hopping, break this self-duality and host energy-dependent mobility edges, multiple re-entrant localization transitions, and mixed-phase regimes~\cite{DasSarma1988,DasSarma1990,Biddle2009,Biddle2010,Biddle2011,Ganeshan2015,Li2020,Roy2021,Yahyavi2019,Roy2022,giriprbl}.

Coupling two quasiperiodic chains into a ladder yields intermediate phases and mobility edges controlled by inter-leg hybridization, asymmetric modulation, or geometry~\cite{Sil2008,Rossignolo2019,Wang2021,Ganguly2023}, and flux-dependent mobility edges have been realized experimentally in synthetic zigzag chains~\cite{An2018}. Such coupled geometries also allow one subsystem to modify the localization of another, providing a natural setting for subsystem-selective and proximity-induced localization (PIL). In interacting systems, a strongly disordered subsystem can localize a weak ergodic bath through the many-body localization proximity effect~\cite{Nandkishore2015,Hyatt2017}, whose interplay with thermalization has been probed with ultracold atoms~\cite{RubioAbadal2019}. At the single-particle level, clean chains coupled to randomly disordered or strongly quasiperiodic chains inherit localization from their strongly disordered partner, tuned by the disorder strength itself~\cite{Zhang2010,Lin2024,Goswami2025}. This raises the question of whether strong disorder is required at all. A conceptually distinct route emerges when the neighboring chain is not strongly disordered but merely detuned: in a ladder of two weakly modulated chains, both extended without detuning, such detuning could imprint localization on the undetuned chain through inter-leg coupling and drive its re-entrant delocalization at larger detuning. Unlike disorder-driven schemes, this would allow transport in one subsystem to be switched reversibly by a staggered detuning of its neighbor alone, a capability relevant to coupled quantum devices and synthetic lattices~\cite{An2018}. This detuning-controlled route to PIL has not yet been explored, and its dynamical signatures have not been observed experimentally.
\begin{figure}[b]
    \centering
    \includegraphics[width=1\columnwidth]{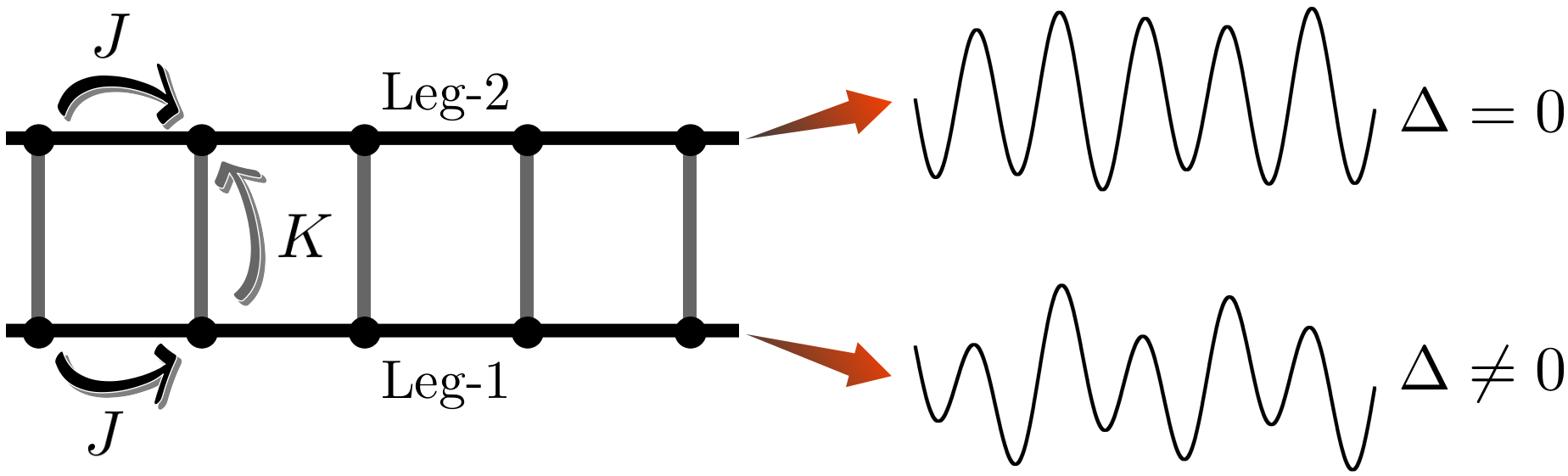}
    \caption{Schematic of the two-leg AA ladder with intra-leg hopping $J$, rung coupling $K$, and a weak quasiperiodic potential on both legs. A staggered detuning $\Delta$ acts only on leg-1, while leg-2 is left undetuned ($\Delta=0$).}
    \label{fig:lattice}
\end{figure}
In this Letter, we demonstrate detuning-controlled PIL in a two-leg AA ladder (Fig.~\ref{fig:lattice}) in which a staggered detuning acts only on leg-1, while leg-2 is left undetuned. Any localization of leg-2 must therefore originate from hybridization with leg-1, cleanly separating proximity-induced from direct localization. We find that PIL emerges only at intermediate inter-leg coupling, where leg-2 begins to localize even before the detuned leg-1. Beyond static diagnostics, PIL and re-entrant delocalization should be resolvable through their dynamical signatures. Complementary to ultracold atoms~\cite{Roati2008,Billy2008} and photonic lattices~\cite{Lahini2009,Schwartz2007}, superconducting quantum processors offer a flexible digital platform in which the Hamiltonian, potential profile, initial state, and observables are reconfigured at the circuit level without modifying the hardware. Resolving localization on such devices is, however, demanding: it emerges only as the evolution time grows, whereas the Trotterized circuit depth grows linearly with time and quickly exceeds what noisy processors execute reliably. Using tensor-network-assisted circuit compression, which reduces the circuit depth and two-qubit gate count by about an order of magnitude, we implement the real-time dynamics of an 18-qubit ladder on IBM Quantum (IBMQ) hardware, measure the leg-resolved wave-packet spreading and participation entropy, and resolve the extended, proximity-localized, and re-entrant regimes.

{\bf{\textit{Model and Approach.-~}}} 
We consider the two-leg AA ladder of Fig.~\ref{fig:lattice}, described by the Hamiltonian $H = H_{1} + H_{2} + H_{12}$, where
\begin{equation}
    \begin{split}
    H_{1} &= -J\sum_{i} \;(\hat{{a}}_{i}^{\dagger} \hat{a}_{i+1} + h.c.) + \lambda \sum_{i} \cos(2\pi\beta i +\phi_{1}) \hat{n}^{a}_{i}\\
    & + \Delta \sum_{i} (-1)^{i}  \hat{n}^{a}_{i}\\
    H_{2} &= -J\sum_{i} \;(\hat{{b}}_{i}^{\dagger} \hat{b}_{i+1} + h.c.) + \lambda \sum_{i} \cos(2\pi\beta i +\phi_{2}) \hat{n}^{b}_{i}\\ 
    H_{12}& = -K\sum_{i}\; (\hat{{a}}_{i}^{\dagger} \hat{b}_{i}  + h.c.)
    \end{split} 
    \label{eqn:ham}
\end{equation}
Here, $H_1$ and $H_2$ describe the staggered AA (s-AA) chain on leg-1 and the pure AA (p-AA) chain on leg-2, respectively, while $H_{12}$ represents the inter-leg coupling, with $\hat{n}^{a}_{i}=\hat{a}_{i}^{\dagger}\hat{a}_{i}$ and $\hat{n}^{b}_{i}=\hat{b}_{i}^{\dagger}\hat{b}_{i}$.
The operators $\hat{a}_i$ ($\hat{b}_i$) and $\hat{a}_i^{\dagger}$ ($\hat{b}_i^{\dagger}$) represent the annihilation and creation operators at site $i$ of leg-1 (leg-2), respectively. The parameters $J$ and $K$ correspond to the intra-leg and inter-leg hopping amplitudes, respectively. Here, $\beta = 2/(\sqrt{5} + 1)$ denotes the inverse golden ratio, $\lambda$ characterizes the strength of the quasiperiodic disorder potential, and $\Delta$ specifies the on-site staggered detuning potential applied only to leg-1. For sufficiently large system sizes, the influence of $\phi_1$ and $\phi_2$ can be neglected as their individual contributions diminish relative to the system's overall behavior. We employ open boundary conditions in the numerical calculations, with $J = 1$ serving as the energy unit. 
%Note that we denote Leg-2 as p-AA, representing a pure AA chain, and Leg-1 as s-AA, representing an AA chain with a staggered on-site potential.

%\section*{Approach}
The transition from extended to localized states can be identified by simultaneously evaluating two key quantities: the inverse participation ratio (IPR) and the normalized participation ratio (NPR) of the Hamiltonian's eigenstates~\cite{DominguezCastro2019}. 
%We focus on the single-particle excitation and express the $m$-th eigenstate as $|\psi_{m}\rangle=\sum_\alpha \psi_{m,\alpha} |\alpha\rangle$, where $|\alpha\rangle$ is the complete single-particle state basis.
For the $m$-th eigenstate, these quantities are defined as $\mathrm{IPR}_{m} = \sum_{\ell=1}^{2}\sum_{i=1}^{L} |\psi_{m}(\ell,i)|^{4}$ and $\mathrm{NPR}_{m} = (N\,\mathrm{IPR}_{m})^{-1}$.
We average them over all eigenstates, $\langle \mathrm{IPR} \rangle= \frac{1}{N}\sum_{m=1}^{N} \mathrm{IPR}_m$ and $\langle \mathrm{NPR} \rangle=\frac{1}{N}\sum_{m=1}^{N} \mathrm{NPR}_m$.
%\begin{equation}
%    \begin{split}
%        \langle IPR \rangle &= \frac{1}{N}\sum_{m=1}^{N} IPR_m\\
%        \langle NPR \rangle &= \frac{1}{N}\sum_{n=1}^{N} NPR_m,
%    \end{split}
%\end{equation}
Here $N=2L$ is the system size and $L$ is the length of each leg.
To identify the intermediate phase (IP), where extended and localized states coexist, we use the parameter $\eta$ defined in Refs.~\cite{Li2020,Roy2021,giriprbl}:
\begin{equation}
    \eta = \log_{10}[\langle \mathrm{IPR} \rangle \times \langle \mathrm{NPR} \rangle].
    \label{eqn:eta}
\end{equation}
%In the extended and localized regions, $\langle \text{IPR} \rangle \sim N^{-1}$ and $\langle \text{NPR} \rangle \sim N^{-1}$, respectively. Consequently, for both extended and localized states, $\eta < -\log_{10}N$. 
%In the extended phase, ($\langle \text{IPR} \rangle \sim N^{-1}$) while ($\langle \text{NPR} \rangle \sim \mathcal{O}(1)$); conversely, in the localized phase, ($\langle \text{IPR} \rangle \sim \mathcal{O}(1)$) and ($\langle \text{IPR} \rangle \sim N^{-1}$). Therefore, in both the extended and localized regimes, the quantity ($\eta$) scales as ($\eta \sim -\log_{10}N$). In our calculations, where $N > 10^3$, this results in $\eta < -3$ for both extended and localized states. In the intermediate region, however, both $\langle \text{IPR} \rangle$ and $\langle \text{NPR} \rangle$ remain finite ($\sim \mathcal{O}(1)$). Thus, in our calculations, the system will exhibit an intermediate phase when $-3 < \eta < -1$.
In the extended phase, $\langle \mathrm{IPR} \rangle \sim N^{-1}$ and $\langle \mathrm{NPR} \rangle \sim \mathcal{O}(1)$, whereas in the localized phase the scaling behavior is reversed. Therefore, in both the extended and localized regimes, $\eta \sim -\log_{10}N$. In our calculations, where $N > 10^3$, this results in $\eta < -3$ for both extended and localized states. In contrast, within the intermediate phase, both $\langle \mathrm{IPR} \rangle$ and $\langle \mathrm{NPR} \rangle$ remain finite $(\sim \mathcal{O}(1))$, leading to $-3 < \eta < -1$. 
\begin{figure}[t]
    \centering    
    \includegraphics[width =1\linewidth]{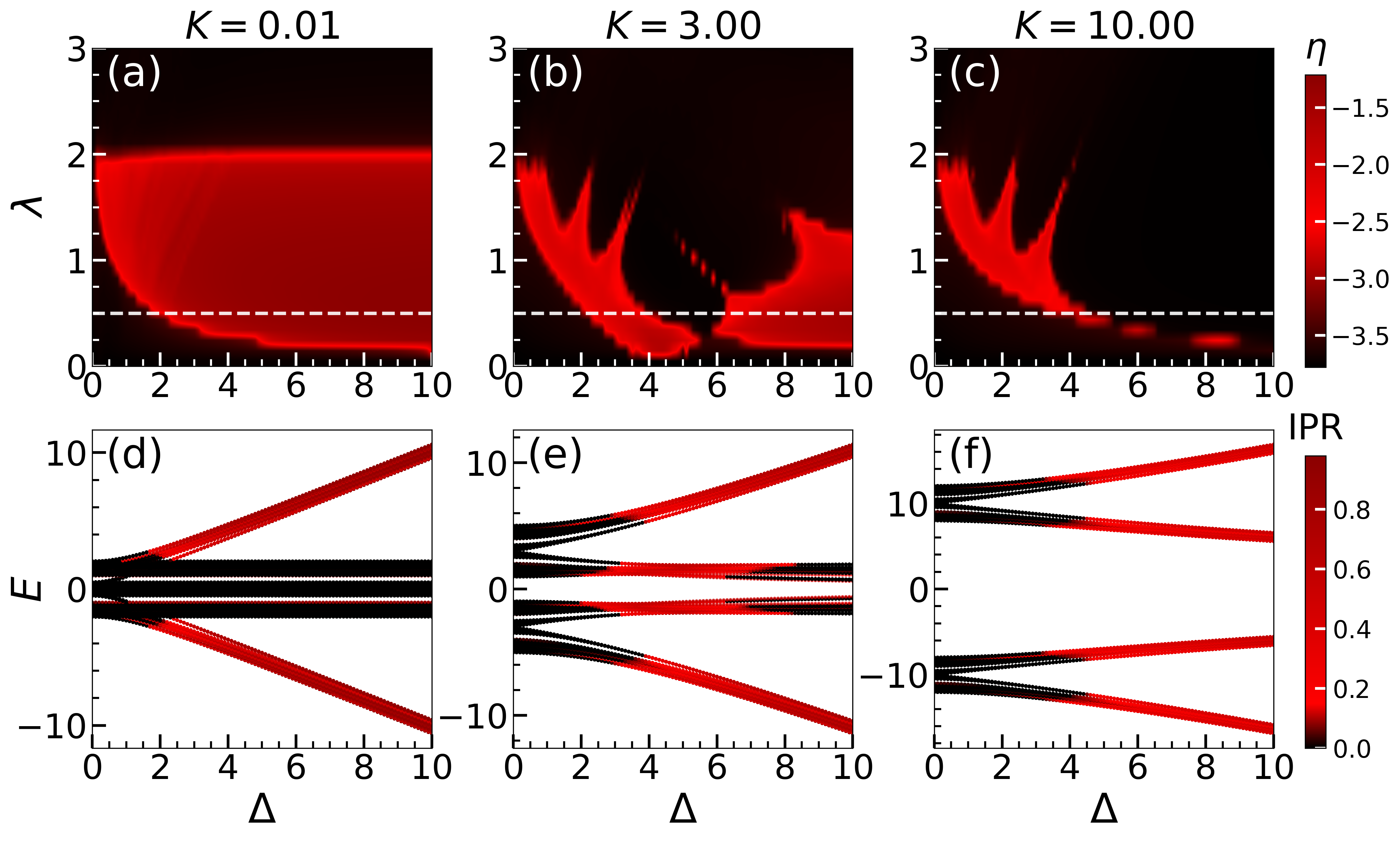}
    \caption{Panels (a)--(c) show the phase diagrams in the $\lambda$--$\Delta$ plane for $K=0.01$, $3.0$, and $10.0$, respectively, where the color scale represents $\eta$ defined in Eq.~(\ref{eqn:eta}); the dashed white line marks $\lambda=0.5$. The black regions represent the extended or localized regimes, while the red regions indicate the intermediate phase (IP), where extended and localized eigenstates coexist. Panels (d)--(f) show the corresponding energy spectra as a function of $\Delta$ at $\lambda=0.5$, with the color scale denoting the inverse participation ratio $\mathrm{IPR}$ of the eigenstate. Here, each leg has $L=3000$ sites.}
    \label{fig:phase_diagram}
\end{figure}
%\section*{Result and Discussion}
{\bf{\textit{Results and Discussion.-}}} 
Figures~\ref{fig:phase_diagram}(a)--(c) show the phase diagram in the $\lambda$--$\Delta$ plane for three representative inter-leg couplings, $K=0.01$, $3.0$, and $10.0$. At $\Delta=0$, the ladder follows the standard AA transition at $\lambda_c=2$, independent of $K$~\cite{PhysRevB.99.054211}. For $\lambda>2$, the quasiperiodic modulation dominates and the states remain localized for the range of $\Delta$ and $K$ considered. We therefore focus on $\lambda=0.5$, marked by the dashed line, where the ladder remains extended at $\Delta=0$. Starting from this extended regime, we examine how tuning the staggered on-site potential $\Delta$ on leg-1 modifies localization on the neighboring leg-2 for different $K$.

%Thus, to observe the proximity effect, we focus mainly on the weak-quasiperiodic modulation strength, $\lambda=0.5$, where the uncoupled legs are extended at $\Delta=0$, and study how tuning the staggered on-site potential $\Delta$ on s-AA leg modifies the localization of the p-AA leg for the inter-leg coupling $K$ strengths.

\begin{figure}[t]
    \centering
    \includegraphics[width =1\columnwidth]{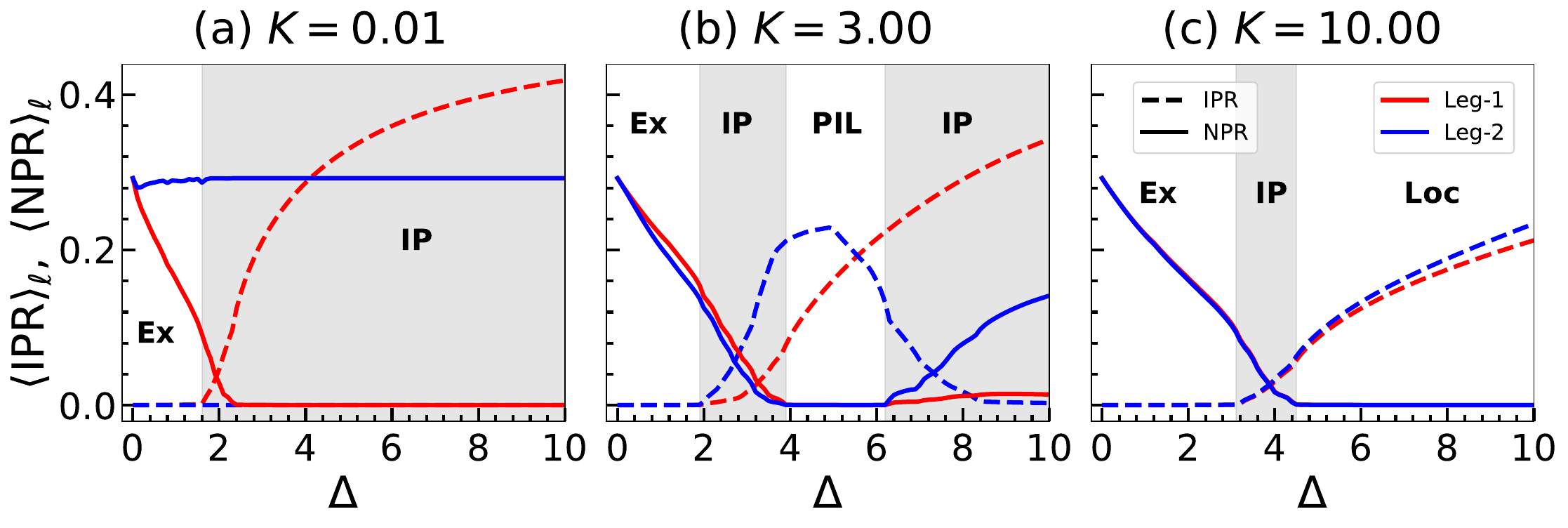}
    \caption{Leg-resolved $\langle \mathrm{IPR} \rangle_{\ell}$ and $\langle \mathrm{NPR} \rangle_{\ell}$ as a function of $\Delta$ for (a) $K=0.01$, (b) $K=3.0$, and (c) $K=10$. Here, each leg has $L=3000$ sites and $\lambda=0.5$. The labels denote Ex: extended phase, IP: intermediate phase, PIL: proximity-induced localization, and Loc: localized phase.}
    \label{fig:IPR_leg_a_b}
\end{figure}

\textit{Weak-coupling limit.-~} For weak inter-leg coupling ($K = 0.01$), the system exhibits a broad IP region for $\lambda \leq 2$, as shown in Fig.~\ref{fig:phase_diagram}(a). This behavior is also reflected in the $E$--$\Delta$ spectrum at $\lambda=0.5$, shown in Fig.~\ref{fig:phase_diagram}(d). In this limit, the two legs are weakly hybridized and therefore largely retain their individual localization properties. The states associated with leg-2 remain extended over the full range of $\Delta$. In contrast, leg-1 acquires larger IPR as $\Delta$ increases, indicating a crossover from extended to localized behavior on the staggered leg. This is further confirmed by the leg-resolved $\langle\mathrm{IPR}\rangle_{\ell}$ and $\langle\mathrm{NPR}\rangle_{\ell}$~\cite{definition} shown in Fig.~\ref{fig:IPR_leg_a_b}(a), where leg-2 remains delocalized while leg-1 undergoes a localization crossover around $\Delta\simeq1.7$. Thus, the weak-coupling IP reflects the coexistence of predominantly localized leg-1 states and extended leg-2 states.

%For weak inter-chain coupling ($K/J = 0.01$), the system exhibits a broad intermediate-phase region for $\lambda \leq 2$, as shown in Fig.~\ref{fig:phase_diagram}(a). This behavior is also reflected in the $E- \Delta$ spectrum at $\lambda=0.5$, shown in Fig.~\ref{fig:phase_diagram}(d). In this limit, the two chains largely preserve their individual localization properties. The p-AA chain (leg-2) remains delocalized over the entire range of $\Delta$, whereas the s-AA chain (leg-1) undergoes a delocalization-to-localization transition with increasing $\Delta$. This can be clearly seen from the leg-resolved $\langle\mathrm{IPR}\rangle_{\ell}$ and $\langle\mathrm{NPR}\rangle_{\ell}$ shown in Fig.~\ref{fig:IPR_leg_a_b}(a).Thus, the intermediate phase in the weak-coupling regime originates from the coexistence of extended states on the leg-2 and localized states on the leg-2. Thus, the coexistence of extended states from the leg-2 and localized states from the leg-1 gives rise to the intermediate phase in the weak-coupling regime. 
%, confirming that the p-AA leg remains extended while the s-AA leg undergoes a delocalization-to-localization crossover with increasing $\Delta$.

\textit{Strong-coupling limit.-~} For strong inter-leg coupling ($K=10.0$), the two legs are strongly hybridized through the rung coupling. As a result, the ladder effectively behaves as a single staggered AA system. As shown in Fig.~\ref{fig:phase_diagram}(c), this regime exhibits a multiple re-entrant localization transition structure similar to that reported in Ref.~\cite{giriprbl}. For $\lambda=0.5$, the system is extended for $\Delta<3$, enters an IP for $3\leq \Delta \leq 4.2$, and becomes localized for $\Delta>4.2$.
This is visible in the $E$--$\Delta$ spectrum in Fig.~\ref{fig:phase_diagram}(f), where both hybridized branches show a similar increase in IPR with increasing $\Delta$, indicating that localization develops collectively across the ladder. This behavior is further confirmed by the leg-resolved $\langle\mathrm{IPR}\rangle_{\ell}$ and  $\langle\mathrm{NPR}\rangle_{\ell}$ in Fig.~\ref{fig:IPR_leg_a_b}(c), where the two legs exhibit nearly similar trends with increasing $\Delta$.
%For strong inter-chain coupling ($K/J = 10.0$), the rungs of the two chains strongly hybridize, and the system effectively maps onto a single staggered AA chain. As shown in Fig.~\ref{fig:phase_diagram}(c), this regime exhibits multiple re-entrant localization transitions with increasing $\Delta$, a signature similar to that reported in Ref~\cite{giriprbl}. For $\lambda=0.5$, the system undergoes a single localization transition with increasing $\Delta$, as reflected in the $E$–$\Delta$ spectrum in Fig.~\ref{fig:phase_diagram}(f), where both bands show a transition from extended to localized states through the intermediate phase. This behavior is quantitatively confirmed in Fig.~\ref{fig:IPR_leg_a_b}(c), where the leg-resolved $\langle \text{IPR} \rangle$ and $\langle \text{NPR} \rangle$ are plotted as a function of $\Delta$, showing that both legs closely follow each other. These results confirm that both chains undergo the same localization transition as the single staggered AA chain reported in Ref~\cite{giriprbl}.
%#################### Write for intermediate coupling  ####################

\textit{Intermediate-coupling limit.-~} For intermediate inter-leg coupling ($K=3.0$), the system exhibits a nontrivial phase structure distinct from both the weak- and strong-coupling limits. As shown in Fig.~\ref{fig:phase_diagram}(b), two separated IP regions appear, with a fully localized regime in between them. This sequence is also clearly reflected in the $E$--$\Delta$ spectrum at $\lambda=0.5$, shown in Fig.~\ref{fig:phase_diagram}(e). For $\Delta < 1.9$, the system is extended. For $1.9 < \Delta < 3.9$, localized and extended states coexist, giving rise to the first IP. In the window $3.9< \Delta < 6.2$, eigenstates acquire large IPR, indicating a fully localized regime. For $\Delta \geq 6.2$, the spectrum enters the second IP, where high-energy states remain localized, while low-energy states recover low IPR, indicating delocalization. Figure~\ref{fig:IPR_leg_a_b}(b) resolves these phases by leg. In the extended region, $\langle\mathrm{IPR}\rangle_\ell$ of both legs vanishes. In the first IP ($1.9<\Delta<3.9$), $\langle\mathrm{IPR}\rangle$ of leg-2 (blue dashed line) rises sharply while that of leg-1 (red dashed line) remains small, indicating that leg-2 starts localizing first despite the staggered detuning applied only to leg-1. Since leg-1 is still nearly extended in this window, the localization of leg-2 cannot be inherited from leg-1 and must instead arise from detuning-induced hybridization between the two legs.
This behavior highlights the proximity-induced nature of localization: although the staggered detuning acts exclusively on leg-1, inter-leg hybridization modifies the effective potential experienced by leg-2 (the p-AA chain), enhancing its localization tendency. With further increasing detuning ($3.9< \Delta < 6.2$), both legs exhibit finite $\langle\mathrm{IPR}\rangle_\ell$ and vanishing $\langle\mathrm{NPR}\rangle_\ell$, identifying the PIL regime.
%This behavior is non-trivial, where rung-induced hybridization reshapes the effective on-site landscape on the leg-2, enhancing the localization tendency of leg-2 relative to leg-1. With further increase in detuning, both legs show finite $\langle \text{IPR} \rangle_\ell$ and vanishing $\langle\mathrm{NPR}\rangle_{\ell}$, marking the PIL regime. 
%In this window, localization on the detuned leg-1 is effectively transferred to the leg-2 through inter-leg hybridization. 
Upon further increasing the detuning ($\Delta \geq 6.2$), the increasing energy separation between the two legs suppresses inter-leg hybridization, progressively reducing the effect of the staggered potential on leg-2. Consequently, leg-2 exhibits a decrease in $\langle\mathrm{IPR}\rangle_\ell$ and a recovery of $\langle\mathrm{NPR}\rangle_\ell$, signaling the onset of re-entrant delocalization on leg-2, while leg-1 remains strongly localized.
%At larger detuning, the system enters a second IP. At larger detuning, the system enters a second IP: leg-2 shows a reduction in $\langle \text{IPR} \rangle_\ell$ together with a recovery of $\langle\mathrm{NPR}\rangle_{\ell}$, signaling re-entrant delocalization, while leg-1 remains strongly localized. 
The site-resolved weight $D(i)$ (\ref{sec:E1_Di}) and a projected-state analysis (\ref{sec:E2_projected}) confirm that this phase structure is robust against the choice of leg-resolved averaging.
%This interpretation is further supported by the site-resolved accumulated weight $D(i)=\sum_n|\psi_n(i)|^4$, presented in the End Matter for representative detuning. 

\begin{figure}[b]
    \centering
    \vspace{-0.2cm}
    \includegraphics[width =1\columnwidth]{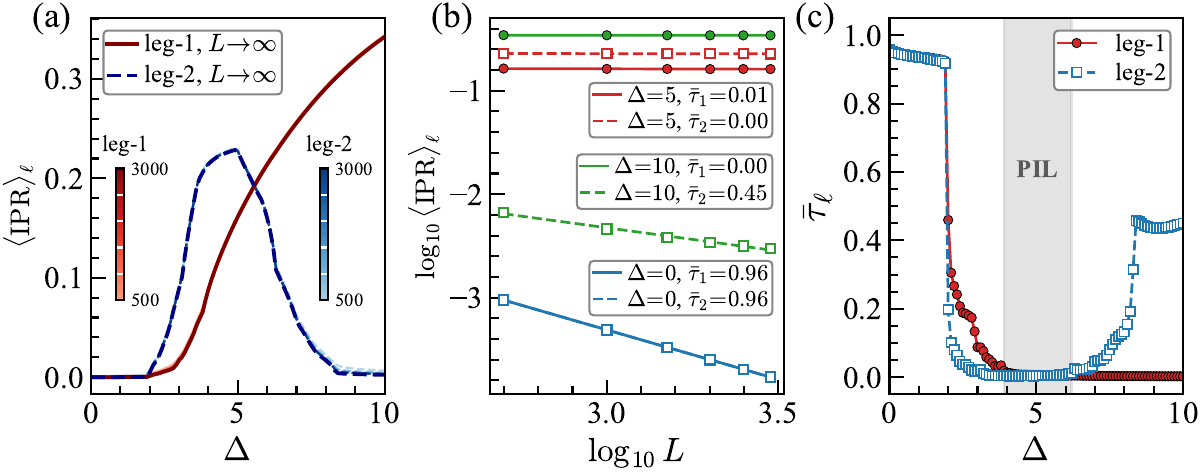}
    \caption{Leg-resolved finite-size scaling for $K=3.0$ and $\lambda=0.5$. (a) $\langle\mathrm{IPR}\rangle_\ell$ versus $\Delta$ for various system sizes $L$ and the extrapolated $L \to \infty$ limit. (b) Extraction of the scaling exponent $\bar{\tau}_\ell$ at representative detuning $\Delta = 0, 5,$ and $10$. (c) Scaling exponent $\bar{\tau}_\ell$ as a function of $\Delta$, highlighting proximity-induced localization ($\bar{\tau}_1 = \bar{\tau}_2 = 0$) and re-entrant delocalization on leg-2 ($\bar{\tau}_2 > 0$).}
%\label{fig:finite_size_scaling}}
    \vspace{-0.5cm}
    \label{fig:fss_d2}
\end{figure}
\textit{Finite-size scaling.-~} 
To verify that the localization features persist toward the thermodynamic limit, we perform finite-size scaling of the leg-resolved average IPR. Figure~\ref{fig:fss_d2}(a) shows $\langle \mathrm{IPR}\rangle_{\ell}$ versus $\Delta$ for different system sizes, together with the $L\rightarrow\infty$ extrapolation obtained from a linear fit in $1/L$. The systematic convergence with increasing system size confirms that the observed localization trends are not finite-size artifacts. In the extrapolated curve, leg-1 follows a single delocalized-to-localized crossover with increasing $\Delta$, whereas leg-2 exhibits a non-monotonic localization behavior. This finite-size analysis therefore supports the proximity-induced localization and re-entrant delocalization of leg-2.

%To verify that the localization features identified above persist toward the thermodynamic limit, we perform a finite-size scaling analysis of the leg-resolved average IPR. Figure~\ref{fig:fss_d2}(a) shows $\langle \mathrm{IPR}\rangle_{\ell}$ as a function of $\Delta$ for different system sizes, together with the extrapolated $L\rightarrow\infty$ curve obtained from a linear fit in $1/L$. The weak size dependence of the curves indicates that the observed localization trends are well converged over the system sizes considered. In the extrapolated curve, the leg-1 shows a gradual increase of $\langle \mathrm{IPR}\rangle_{\ell}$ with increasing $\Delta$, consistent with localization driven by the staggered detuning. In contrast, the leg-2 displays a strongly non-monotonic response, where it first develops a finite $\langle \mathrm{IPR}\rangle_{\ell}$ through inter-leg hybridization, becomes localized together with the leg-1 in the PIL window, and then decreases again at larger $\Delta$. This behavior is consistent with proximity-induced localization followed by re-entrant delocalization on the p-AA leg.

To classify the leg-resolved localization character, we extract the average scaling exponent $\bar{\tau}_{\ell}$ from $\langle \mathrm{IPR}\rangle_{\ell}\sim L^{-\bar{\tau}_{\ell}}$~\cite{Roy2018multifractality,Roy2022,RevModPhys.80.1355,PhysRevLett.123.070405}, obtained from the slope of $\log_{10}\langle \mathrm{IPR}\rangle_{\ell}$ versus $\log_{10}L$ and shown in Fig.~\ref{fig:fss_d2}(b). Here, $\bar{\tau}_{\ell}\sim1$ corresponds to extended behavior, $\bar{\tau}_{\ell}\sim0$ to localized behavior, and intermediate values $0<\bar{\tau}_{\ell}<1$ indicate a mixed localization regime. 
%At $\Delta=0$, both legs yield $\tau_{\ell}=0.96$ [Fig.~\ref{fig:fss_d2}(b)], close to the ideal extended value $\tau_{\ell}=1$. %This confirms an extended regime; the minor deviation from unity stems strictly from quasiperiodic commensurability effects rather than finite-size limitations (see Supplementary).
At $\Delta=0$, both legs give $\bar{\tau}_{\ell}\sim1$ [Fig.~\ref{fig:fss_d2}(b)], consistent with extended behavior. In the PIL regime, represented by $\Delta=5.0$, both legs yield $\bar{\tau}_{\ell}=0$, confirming simultaneous localization across the ladder. At larger detuning, $\Delta=10.0$, the scaling becomes strongly leg asymmetric: leg-1 remains localized with $\bar{\tau}_1=0$, whereas leg-2 recovers a finite exponent $\bar{\tau}_2=0.45$. This intermediate exponent reflects the coexistence of extended-like and localized eigenstates on leg-2 (see \ref{sec:mixed_loc}), providing further evidence of its re-entrant delocalized character.
%At larger detuning, $\Delta=10.0$, the scaling becomes strongly leg asymmetric: leg-1 remains localized with $\bar{\tau}_{1}=0$, whereas leg-2 recovers a finite exponent $\bar{\tau}_{2}=0.45$. This finite exponent indicates a mixed localization character on leg-2 (see EM3) and captures its re-entrant delocalized behavior.

The full $\Delta$ dependence in Fig.~\ref{fig:fss_d2}(c) summarizes this sequence: $\bar{\tau}_2$ drops sharply near $\Delta\simeq2$ and approaches zero well before $\bar{\tau}_1$, directly showing that leg-2 localizes before the detuned leg-1; both exponents then vanish for $3.9<\Delta<6.2$, and only $\bar{\tau}_2$ becomes finite at larger $\Delta$.

%The full $\Delta$-dependence of $\tau_{\ell}$, shown in Fig.~\ref{fig:fss_d2}(c), provides a complete view of this transition sequence. With increasing $\Delta$, $\tau_2$ decreases rapidly, showing that the leg-2 is driven toward localization through hybridization with the detuned leg-1. In the interval $4\leq\Delta\leq6.2$, both exponents remain close to zero, confirming the PIL regime where both legs are localized.For larger $\Delta$, the two legs respond differently. While $\tau_1$ remains close to zero, indicating that the s-AA leg stays localized in the thermodynamic limit, $\tau_2$ increases again to a finite value. This demonstrates the re-entrant recovery of weakly delocalized behavior on the leg-2. Therefore, the large-$\Delta$ regime is an asymmetric intermediate phase in which a localized leg-1 coexists with a partially delocalized leg-2.

\textit{\textbf{Quantum simulation on IBMQ hardware.-~}}
%Having established the localization regimes from the static analysis, we now examine their dynamical signatures. This provides a direct physical probe of PIL: localization is reflected in the suppression of wave-packet spreading, whereas the re-entrant regime is identified by the recovery of transport on the proximity-localized p-AA chain. Superconducting quantum processors offer a flexible digital platform for this purpose, since the Hamiltonian parameters, quasiperiodic profile, initial state, and measured observables can be reconfigured at the circuit level without modifying the underlying hardware. We implement the real-time dynamics on IBMQ by evolving an initial state $|\psi(0)\rangle$ under the time-independent Hamiltonian $H$ defined in Eq.~(\ref{eqn:ham}), $|\psi(t)\rangle = e^{-iHt}|\psi(0)\rangle$, where $e^{-iHt}$ is the time-evolution operator. 
Having established the localization regimes through static analysis, we now probe them experimentally through their real-time dynamics on IBMQ hardware, using exact calculations as a reference. We simulate $|\psi(t)\rangle=e^{-iHt}|\psi(0)\rangle$ under the Hamiltonian in Eq.~(\ref{eqn:ham}), starting from a particle prepared in an equal superposition across the central rung,
$|\psi(0)\rangle=\frac{1}{\sqrt{2}}\left(a^\dagger_{i_0}+b^\dagger_{i_0}\right)|\mathrm{vac}\rangle$,
where $i_0=4$ is the central rung (rungs labeled $i=0,\ldots,L-1$). The site occupations are obtained from local measurements as $n_\ell(i,t)=[1-\langle\hat Z_{\ell,i}\rangle_t]/2$, with $\ell=1,2$ labeling the two legs. From these occupations, we extract the leg-resolved mean-square displacement (MSD)~\cite{hiramoto1988dynamics2,DominguezCastro2019,giriprbl},
$R_\ell^2(t)=\sum_i(i-i_0)^2n_\ell(i,t)$.
Its evolution at different staggered potentials $\Delta$ distinguishes the dynamics in the extended, PIL, and re-entrant delocalized regimes.

%The particle is initially prepared on the central rung of the ladder as $ |\psi(0)\rangle = \frac{1}{\sqrt{2}} \left(a^{\dagger}_{i_0}|0\rangle + b^{\dagger}_{i_0}|0\rangle \right),$ where $i_0=L/2$ denotes the central rung. This state corresponds to a particle equally delocalized between the two legs of the same rung. We compute the local on-site density, $n_{\ell}(i,t) =\frac{1-\langle\hat Z_{\ell,i}\rangle}{2}$,where $\ell=1,2$ labels the two legs. We then compute the leg-resolved mean-square displacement (MSD)~\cite{hiramoto1988dynamics2,DominguezCastro2019,giriprbl}, $R^2_{\ell}(t)= \sum_i (i-i_0)^2 n_{\ell}(i,t)$, for different values of the staggered potential $\Delta$. The MSD directly probes the spreading of the particle wave function and therefore provides a dynamical signature of the extended, PIL, and re-entrant delocalized regimes.
For the $N=18$-qubit ladder, the time-evolution operator is approximated using a first-order Suzuki-Trotter decomposition with time step $\delta t=0.1J^{-1}$. To obtain hardware-executable circuits, we employ a tensor-network-assisted variational approach to compress the Trotterized quantum circuits. The resulting compressed circuits are executed on the IBM Quantum processor \texttt{ibm\_boston} with dynamical decoupling and Pauli twirling, followed by post-selection onto the conserved particle sector, and retain an average hardware Hellinger fidelity of $\bar{\mathcal{F}}\approx0.95$ over all detunings and evolution times studied. The compression fidelity, reduction in transpiled circuit resources and shot-count convergence are reported in \ref{sec:E4_ibmq_implementation} and \ref{sec:E6_shots_study}. Complete details of the circuit construction, compression procedure, and hardware implementation are provided in the SM~\cite{SuppMat}.
%The time-evolution operator, $e^{-iHt}$ is implemented using a Trotterized circuit with time step $\delta t=0.1$, followed by tensor-network-based variational compression to reduce the circuit depth and two-qubit-gate count to do simulation on present day NISQ hardware~\cite{write why we need circuit compression}. The compressed circuits are executed on the IBMQ backend \texttt{ibm\_boston} and benchmarked against exact diagonalization (ED); further implementation details and long-time dynamics from ED are given in the Supplemental Material.
\begin{figure}[t]
    \centering
    %\vspace{0.15cm}
    \includegraphics[width =1\columnwidth]{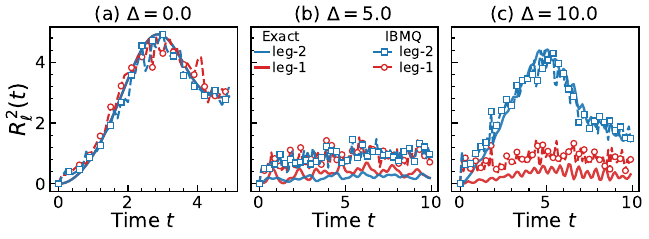}
    \caption{Leg-resolved mean-square displacement $R_\ell^2(t)$ obtained from exact evolution (solid lines) and IBMQ data (dashed lines) for (a) $\Delta=0$, (b) $\Delta=5$, and (c) $\Delta=10$ at $K=3$. Red and blue curves correspond to leg-1 and leg-2, respectively.}
    %\vspace{-0.5cm}
    \label{fig:msd}
\end{figure}
Figures~\ref{fig:msd}(a)--(c) show the leg-resolved MSD, $R^2_{\ell}(t)$, for $\Delta=0$, $5$, and $10$, respectively, obtained from IBMQ and compared with exact time evolution. For $\Delta=0$, both legs exhibit ballistic wave-packet spreading with nearly identical dynamics, consistent with the extended regime identified from the static analysis. The downturn of $R^2_{\ell}(t)$ at later times arises from finite-size boundary reflection after the wave function reaches the edge and should not be interpreted as localization. At $\Delta=5$, corresponding to the PIL regime, the spreading is strongly suppressed on both legs and $R^2_{\ell}(t)$ remains small throughout the evolution. The IBMQ data qualitatively reproduce this suppression, providing hardware-level dynamical evidence of direct localization on leg-1 and proximity-induced localization on leg-2. At $\Delta=10$, the dynamics becomes leg asymmetric. The MSD on leg-1 remains strongly suppressed, whereas leg-2 exhibits a re-entrant recovery of wave function spreading. This selective recovery of wave function spreading is consistent with the finite-size scaling analysis, where leg-1 remains localized while leg-2 recovers a finite scaling exponent, indicating re-entrant delocalized character.
%Thus, the IBMQ results provide an independent dynamical confirmation of the static phase structure: ballistic spreading, strongly suppressed spreading in the PIL regime and restored diffusive transport on the leg-2 at larger detuning (see Supplementary for long-time study).
%Fig.~\ref{fig:msd}(a-c) shows the leg-resolved MSD, $R^2_{\ell}(t)$ measured on IBMQ for $\Delta=0$, $5$, and $10$, respectively.
%The IBMQ results are benchmarked against exact time evolution. 

To further probe the localization dynamics, we examine the experimentally accessible second-order participation entropy on leg-2~\cite{PhysRevLett.112.057203,li2023observation,Giri_2025}, defined as
%The spatial support of the evolving wave function provides a direct dynamical measure of proximity-induced localization. We therefore measure experimental accessible characterization, the second-order dynamical participation entropy on the pure AA chain~\cite{PhysRevLett.112.057203,li2023observation,Giri_2025}, defined as,
\begin{figure}[b]
    \centering
    \includegraphics[width =1\columnwidth]{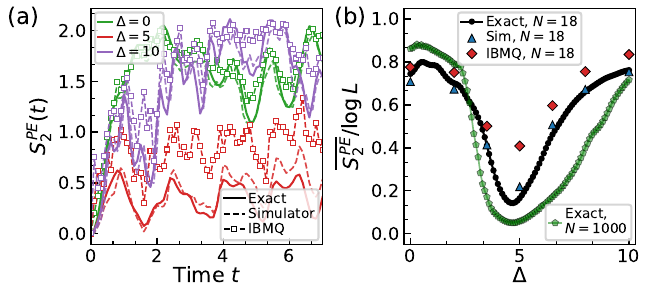}
    \caption{(a) Time evolution of $S^{PE}_{2}$ for $N=18$ at $\Delta=0,5,$ and $10$, obtained from exact dynamics (solid lines), the noiseless simulator (dashed lines), and IBMQ (dashed lines with squares). (b) Time-averaged normalized participation entropy, $\overline{S_{\mathrm{2}}^{PE}}/\log L$, as a function of $\Delta$ for $L=9$ ($N=18$; exact, simulator, and IBMQ) and $L=500$ ($N=1000$; exact). The averages are taken over $t\in[2,7]$ for $L=9$ and $t\in[10^3,1.5\times10^3]$ for $L=500$.}
    \label{fig:PE}
\end{figure}
$S_{\mathrm{2}}^{PE}(t) = -\log\sum_i p_{2,i}^2(t)$. Here, $p_{2,i}(t)=n_2(i,t)/\sum_j n_2(j,t)$ denotes the normalized site-occupation probability on leg-2. This quantity is the negative logarithm of the IPR on leg-2 and it characterizes how fast the wave function spreads on leg-2. Figure~\ref{fig:PE}(a) shows that $S_{\mathrm{2}}^{PE}(t)$ grows rapidly for $\Delta=0$, remains strongly suppressed near $\Delta=5$, and increases again for $\Delta=10$, resolving the extended, proximity-localized, and re-entrant delocalized regimes in real time. The IBMQ data closely follow the exact $L=9$ dynamics over the accessible time window. Note that, for $\Delta=5$, $S_{\mathrm{2}}^{PE}(t)$ shows a larger deviation from the exact dynamics due to enhanced Trotter sensitivity at intermediate detuning (see \ref{sec:E5_error_analysis}). Next, we use the late-time average of participation entropy, $\overline{S_{\mathrm{2}}^{PE}}$, to quantify the detuning-dependent PIL response. This averaged quantity gives a single measure of the effective number of leg-2 sites occupied after the initial transient dynamics. We rescale it as $\overline{S_{\mathrm{2}}^{PE}}/\log L$ to study different system sizes. This normalized quantity approaches unity when the wave function is delocalized over most of leg-2, whereas it remains small when the wave function is confined to only a few sites. Figure~\ref{fig:PE}(b) shows $\overline{S_{\mathrm{2}}^{PE}}/\log L$ as a function of $\Delta$. For $L=9$, the IBMQ data at $\Delta=0, 2.0, 3.5, 5.0, 6.5, 8.0$, and $10.0$ are compared with exact dynamics at the same system size, showing that the digital quantum simulation captures the transition from a delocalized to a PIL regime and the subsequent re-entrant delocalized regime on leg-2. We further include exact dynamics for a larger system with $L=500$, averaged over the long-time window $t\in[10^3,1.5\times10^3]$ to capture its long-time behavior, to assess finite-size effects beyond the quantum-device size. Compared with $L=9$, the $L=500$ result exhibits a sharper minimum in the PIL regime, indicating that the suppression of participation entropy persists at larger system sizes.
%Compared with $L=9$, the $L=500$ result shows a sharper minimum of $\overline{S_{\mathrm{2}}^{PE}}$ in the PIL regime, confirming that the entropy minimum is not a small-system artifact. This establishes the detuning dependence of $\overline{S_{\mathrm{2}}^{PE}}$ as a robust dynamical indicator of the leg-2 localization crossover.

\textit{\textbf{Conclusion.-~}} We have demonstrated detuning-controlled proximity-induced localization and re-entrant delocalization in a weakly quasiperiodic two-leg ladder, without any strong disorder. At intermediate inter-leg coupling, a staggered detuning applied only to leg-1 localizes the undetuned leg-2 through inter-leg hybridization, remarkably before leg-1 itself localizes, establishing that the localization of leg-2 is not inherited from leg-1. At larger detuning, the increasing energy separation between the legs suppresses hybridization: leg-1 remains localized, whereas leg-2 re-enters a delocalized regime, as confirmed by leg-resolved diagnostics and finite-size scaling. Using tensor-network-assisted circuit compression on IBMQ hardware, we observed the corresponding suppression and recovery of wave-packet spreading.
%Our quantum simulation using a QPU provides an independent, operational validation of this phase diagram. Both exact calculations and hardware-based time evolution show a clear evolution from ballistic spreading in the clean limit, to frozen transport within the PIL window, and finally to a re-entrant recovery of transport restricted to the p-AA chain or leg-2 at large detuning. 
%These results establish detuning as a controllable mechanism for selectively inducing and releasing localization in coupled systems. Looking forward, exploring how many-body interactions affect these leg-selective phases presents an exciting future direction, while our current work provides an immediately applicable blueprint for engineering asymmetric transport on programmable quantum simulators.
These results establish leg-selective detuning as a reversible control knob for localization and transport in coupled quantum systems, directly applicable to synthetic-lattice and photonic ladders, and open the way to exploring how many-body interactions shape such proximity effects.

\textit{Acknowledgment.-~} MKG thanks Ranjan Modak for fruitful discussions. We acknowledge the IBM Quantum Hub at National Taiwan University for providing access to IBM Quantum devices. The authors acknowledge using OpenAI's ChatGPT 5.6 and Anthropic's Claude to make minor improvements to the presentation of figures and for language editing; all suggestions were reviewed and approved by the authors. This work was supported by the National Science and Technology Council (NSTC) of Taiwan through Grant No. 115-2119-M-007-005. PS and MKG acknowledge the support of the Singapore Ministry of Education Academic Research Fund's Tier 3 Grant MOE-MOET32023-0003 (PS) and Tier 2 Grant MOE-T2EP50123-0021 (MKG). PC acknowledges the support by Taiwan Centers of Excellence (TCE), Ministry of Education, Taiwan.

\bibliography{PR.bib}

\begin{thebibliography}{53}
\expandafter\ifx\csname natexlab\endcsname\relax\def\natexlab#1{#1}\fi
\expandafter\ifx\csname bibnamefont\endcsname\relax
  \def\bibnamefont#1{#1}\fi
\expandafter\ifx\csname bibfnamefont\endcsname\relax
  \def\bibfnamefont#1{#1}\fi
\expandafter\ifx\csname citenamefont\endcsname\relax
  \def\citenamefont#1{#1}\fi
\expandafter\ifx\csname url\endcsname\relax
  \def\url#1{\texttt{#1}}\fi
\expandafter\ifx\csname urlprefix\endcsname\relax\def\urlprefix{URL }\fi
\providecommand{\bibinfo}[2]{#2}
\providecommand{\eprint}[2][]{\url{#2}}

\bibitem[{\citenamefont{Anderson}(1958)}]{Anderson1958}
\bibinfo{author}{\bibfnamefont{P.~W.} \bibnamefont{Anderson}}, \bibinfo{journal}{Phys. Rev.} \textbf{\bibinfo{volume}{109}}, \bibinfo{pages}{1492} (\bibinfo{year}{1958}).

\bibitem[{\citenamefont{Abrahams et~al.}(1979)\citenamefont{Abrahams, Anderson, Licciardello, and Ramakrishnan}}]{Abrahams1979}
\bibinfo{author}{\bibfnamefont{E.}~\bibnamefont{Abrahams}}, \bibinfo{author}{\bibfnamefont{P.~W.} \bibnamefont{Anderson}}, \bibinfo{author}{\bibfnamefont{D.~C.} \bibnamefont{Licciardello}}, \bibnamefont{and} \bibinfo{author}{\bibfnamefont{T.~V.} \bibnamefont{Ramakrishnan}}, \bibinfo{journal}{Phys. Rev. Lett.} \textbf{\bibinfo{volume}{42}}, \bibinfo{pages}{673} (\bibinfo{year}{1979}).

\bibitem[{\citenamefont{Lee and Ramakrishnan}(1985)}]{LeeRamakrishnan1985}
\bibinfo{author}{\bibfnamefont{P.~A.} \bibnamefont{Lee}} \bibnamefont{and} \bibinfo{author}{\bibfnamefont{T.~V.} \bibnamefont{Ramakrishnan}}, \bibinfo{journal}{Rev. Mod. Phys.} \textbf{\bibinfo{volume}{57}}, \bibinfo{pages}{287} (\bibinfo{year}{1985}).

\bibitem[{\citenamefont{Evers and Mirlin}(2008)}]{RevModPhys.80.1355}
\bibinfo{author}{\bibfnamefont{F.}~\bibnamefont{Evers}} \bibnamefont{and} \bibinfo{author}{\bibfnamefont{A.~D.} \bibnamefont{Mirlin}}, \bibinfo{journal}{Rev. Mod. Phys.} \textbf{\bibinfo{volume}{80}}, \bibinfo{pages}{1355} (\bibinfo{year}{2008}).

\bibitem[{\citenamefont{Soukoulis and Economou}(1982)}]{Soukoulis1982}
\bibinfo{author}{\bibfnamefont{C.~M.} \bibnamefont{Soukoulis}} \bibnamefont{and} \bibinfo{author}{\bibfnamefont{E.~N.} \bibnamefont{Economou}}, \bibinfo{journal}{Phys. Rev. Lett.} \textbf{\bibinfo{volume}{48}}, \bibinfo{pages}{1043} (\bibinfo{year}{1982}).

\bibitem[{\citenamefont{Das~Sarma et~al.}(1988{\natexlab{a}})\citenamefont{Das~Sarma, He, and Xie}}]{PhysRevLett.61.2144}
\bibinfo{author}{\bibfnamefont{S.}~\bibnamefont{Das~Sarma}}, \bibinfo{author}{\bibfnamefont{S.}~\bibnamefont{He}}, \bibnamefont{and} \bibinfo{author}{\bibfnamefont{X.~C.} \bibnamefont{Xie}}, \bibinfo{journal}{Physical Review Letters} \textbf{\bibinfo{volume}{61}}, \bibinfo{pages}{2144} (\bibinfo{year}{1988}{\natexlab{a}}).

\bibitem[{\citenamefont{Lellouch and Sanchez-Palencia}(2014)}]{PhysRevA.90.061602}
\bibinfo{author}{\bibfnamefont{S.}~\bibnamefont{Lellouch}} \bibnamefont{and} \bibinfo{author}{\bibfnamefont{L.}~\bibnamefont{Sanchez-Palencia}}, \bibinfo{journal}{Physical Review A} \textbf{\bibinfo{volume}{90}}, \bibinfo{pages}{061602(R)} (\bibinfo{year}{2014}).

\bibitem[{\citenamefont{Boers et~al.}(2007)\citenamefont{Boers, Goedeke, Hinrichs, and Holthaus}}]{PhysRevA.75.063404}
\bibinfo{author}{\bibfnamefont{D.~J.} \bibnamefont{Boers}}, \bibinfo{author}{\bibfnamefont{B.}~\bibnamefont{Goedeke}}, \bibinfo{author}{\bibfnamefont{D.}~\bibnamefont{Hinrichs}}, \bibnamefont{and} \bibinfo{author}{\bibfnamefont{M.}~\bibnamefont{Holthaus}}, \bibinfo{journal}{Physical Review A} \textbf{\bibinfo{volume}{75}}, \bibinfo{pages}{063404} (\bibinfo{year}{2007}).

\bibitem[{\citenamefont{Biddle and Das~Sarma}(2010)}]{Biddle2010}
\bibinfo{author}{\bibfnamefont{J.}~\bibnamefont{Biddle}} \bibnamefont{and} \bibinfo{author}{\bibfnamefont{S.}~\bibnamefont{Das~Sarma}}, \bibinfo{journal}{Physical Review Letters} \textbf{\bibinfo{volume}{104}}, \bibinfo{pages}{070601} (\bibinfo{year}{2010}).

\bibitem[{\citenamefont{Ganeshan et~al.}(2015)\citenamefont{Ganeshan, Pixley, and Das~Sarma}}]{Ganeshan2015}
\bibinfo{author}{\bibfnamefont{S.}~\bibnamefont{Ganeshan}}, \bibinfo{author}{\bibfnamefont{J.~H.} \bibnamefont{Pixley}}, \bibnamefont{and} \bibinfo{author}{\bibfnamefont{S.}~\bibnamefont{Das~Sarma}}, \bibinfo{journal}{Physical Review Letters} \textbf{\bibinfo{volume}{114}}, \bibinfo{pages}{146601} (\bibinfo{year}{2015}).

\bibitem[{\citenamefont{Aubry and Andr{\'e}}(1980)}]{AubryAndre1980}
\bibinfo{author}{\bibfnamefont{S.}~\bibnamefont{Aubry}} \bibnamefont{and} \bibinfo{author}{\bibfnamefont{G.}~\bibnamefont{Andr{\'e}}}, \bibinfo{journal}{Annals of the Israel Physical Society} \textbf{\bibinfo{volume}{3}}, \bibinfo{pages}{133} (\bibinfo{year}{1980}).

\bibitem[{\citenamefont{Aulbach et~al.}(2004)\citenamefont{Aulbach, Wobst, Ingold, H{\"a}nggi, and Varga}}]{Aulbach_2004}
\bibinfo{author}{\bibfnamefont{C.}~\bibnamefont{Aulbach}}, \bibinfo{author}{\bibfnamefont{A.}~\bibnamefont{Wobst}}, \bibinfo{author}{\bibfnamefont{G.-L.} \bibnamefont{Ingold}}, \bibinfo{author}{\bibfnamefont{P.}~\bibnamefont{H{\"a}nggi}}, \bibnamefont{and} \bibinfo{author}{\bibfnamefont{I.}~\bibnamefont{Varga}}, \bibinfo{journal}{New J. Phys.} \textbf{\bibinfo{volume}{6}}, \bibinfo{pages}{70} (\bibinfo{year}{2004}).

\bibitem[{\citenamefont{Das~Sarma et~al.}(1988{\natexlab{b}})\citenamefont{Das~Sarma, He, and Xie}}]{DasSarma1988}
\bibinfo{author}{\bibfnamefont{S.}~\bibnamefont{Das~Sarma}}, \bibinfo{author}{\bibfnamefont{S.}~\bibnamefont{He}}, \bibnamefont{and} \bibinfo{author}{\bibfnamefont{X.~C.} \bibnamefont{Xie}}, \bibinfo{journal}{Physical Review Letters} \textbf{\bibinfo{volume}{61}}, \bibinfo{pages}{2144} (\bibinfo{year}{1988}{\natexlab{b}}).

\bibitem[{\citenamefont{Das~Sarma et~al.}(1990)\citenamefont{Das~Sarma, He, and Xie}}]{DasSarma1990}
\bibinfo{author}{\bibfnamefont{S.}~\bibnamefont{Das~Sarma}}, \bibinfo{author}{\bibfnamefont{S.}~\bibnamefont{He}}, \bibnamefont{and} \bibinfo{author}{\bibfnamefont{X.~C.} \bibnamefont{Xie}}, \bibinfo{journal}{Physical Review B} \textbf{\bibinfo{volume}{41}}, \bibinfo{pages}{5544} (\bibinfo{year}{1990}).

\bibitem[{\citenamefont{Biddle et~al.}(2009)\citenamefont{Biddle, Wang, Priour, and Das~Sarma}}]{Biddle2009}
\bibinfo{author}{\bibfnamefont{J.}~\bibnamefont{Biddle}}, \bibinfo{author}{\bibfnamefont{B.}~\bibnamefont{Wang}}, \bibinfo{author}{\bibfnamefont{D.~J.} \bibnamefont{Priour}}, \bibnamefont{and} \bibinfo{author}{\bibfnamefont{S.}~\bibnamefont{Das~Sarma}}, \bibinfo{journal}{Physical Review A} \textbf{\bibinfo{volume}{80}}, \bibinfo{pages}{021603} (\bibinfo{year}{2009}).

\bibitem[{\citenamefont{Biddle et~al.}(2011)\citenamefont{Biddle, Priour, Wang, and Das~Sarma}}]{Biddle2011}
\bibinfo{author}{\bibfnamefont{J.}~\bibnamefont{Biddle}}, \bibinfo{author}{\bibfnamefont{D.~J.} \bibnamefont{Priour}}, \bibinfo{author}{\bibfnamefont{B.}~\bibnamefont{Wang}}, \bibnamefont{and} \bibinfo{author}{\bibfnamefont{S.}~\bibnamefont{Das~Sarma}}, \bibinfo{journal}{Physical Review B} \textbf{\bibinfo{volume}{83}}, \bibinfo{pages}{075105} (\bibinfo{year}{2011}).

\bibitem[{\citenamefont{Li and Das~Sarma}(2020)}]{Li2020}
\bibinfo{author}{\bibfnamefont{X.}~\bibnamefont{Li}} \bibnamefont{and} \bibinfo{author}{\bibfnamefont{S.}~\bibnamefont{Das~Sarma}}, \bibinfo{journal}{Physical Review B} \textbf{\bibinfo{volume}{101}}, \bibinfo{pages}{064203} (\bibinfo{year}{2020}).

\bibitem[{\citenamefont{Roy et~al.}(2021)\citenamefont{Roy, Mishra, Tanatar, and Basu}}]{Roy2021}
\bibinfo{author}{\bibfnamefont{S.}~\bibnamefont{Roy}}, \bibinfo{author}{\bibfnamefont{T.}~\bibnamefont{Mishra}}, \bibinfo{author}{\bibfnamefont{B.}~\bibnamefont{Tanatar}}, \bibnamefont{and} \bibinfo{author}{\bibfnamefont{S.}~\bibnamefont{Basu}}, \bibinfo{journal}{Physical Review Letters} \textbf{\bibinfo{volume}{126}}, \bibinfo{pages}{106803} (\bibinfo{year}{2021}).

\bibitem[{\citenamefont{Yahyavi et~al.}(2019)\citenamefont{Yahyavi, Het{\'e}nyi, and Tanatar}}]{Yahyavi2019}
\bibinfo{author}{\bibfnamefont{M.}~\bibnamefont{Yahyavi}}, \bibinfo{author}{\bibfnamefont{B.}~\bibnamefont{Het{\'e}nyi}}, \bibnamefont{and} \bibinfo{author}{\bibfnamefont{B.}~\bibnamefont{Tanatar}}, \bibinfo{journal}{Physical Review B} \textbf{\bibinfo{volume}{100}}, \bibinfo{pages}{064202} (\bibinfo{year}{2019}).

\bibitem[{\citenamefont{Roy et~al.}(2022)\citenamefont{Roy, Chattopadhyay, Mishra, and Basu}}]{Roy2022}
\bibinfo{author}{\bibfnamefont{S.}~\bibnamefont{Roy}}, \bibinfo{author}{\bibfnamefont{S.}~\bibnamefont{Chattopadhyay}}, \bibinfo{author}{\bibfnamefont{T.}~\bibnamefont{Mishra}}, \bibnamefont{and} \bibinfo{author}{\bibfnamefont{S.}~\bibnamefont{Basu}}, \bibinfo{journal}{Physical Review B} \textbf{\bibinfo{volume}{105}}, \bibinfo{pages}{214203} (\bibinfo{year}{2022}).

\bibitem[{\citenamefont{Padhan et~al.}(2022)\citenamefont{Padhan, Giri, Mondal, and Mishra}}]{giriprbl}
\bibinfo{author}{\bibfnamefont{A.}~\bibnamefont{Padhan}}, \bibinfo{author}{\bibfnamefont{M.~K.} \bibnamefont{Giri}}, \bibinfo{author}{\bibfnamefont{S.}~\bibnamefont{Mondal}}, \bibnamefont{and} \bibinfo{author}{\bibfnamefont{T.}~\bibnamefont{Mishra}}, \bibinfo{journal}{Phys. Rev. B} \textbf{\bibinfo{volume}{105}}, \bibinfo{pages}{L220201} (\bibinfo{year}{2022}).

\bibitem[{\citenamefont{Sil et~al.}(2008)\citenamefont{Sil, Maiti, and Chakrabarti}}]{Sil2008}
\bibinfo{author}{\bibfnamefont{S.}~\bibnamefont{Sil}}, \bibinfo{author}{\bibfnamefont{S.~K.} \bibnamefont{Maiti}}, \bibnamefont{and} \bibinfo{author}{\bibfnamefont{A.}~\bibnamefont{Chakrabarti}}, \bibinfo{journal}{Physical Review Letters} \textbf{\bibinfo{volume}{101}}, \bibinfo{pages}{076803} (\bibinfo{year}{2008}).

\bibitem[{\citenamefont{Rossignolo and Dell'Anna}(2019{\natexlab{a}})}]{Rossignolo2019}
\bibinfo{author}{\bibfnamefont{M.}~\bibnamefont{Rossignolo}} \bibnamefont{and} \bibinfo{author}{\bibfnamefont{L.}~\bibnamefont{Dell'Anna}}, \bibinfo{journal}{Physical Review B} \textbf{\bibinfo{volume}{99}}, \bibinfo{pages}{054211} (\bibinfo{year}{2019}{\natexlab{a}}).

\bibitem[{\citenamefont{Wang et~al.}(2021)\citenamefont{Wang, Yang, and Song}}]{Wang2021}
\bibinfo{author}{\bibfnamefont{R.}~\bibnamefont{Wang}}, \bibinfo{author}{\bibfnamefont{X.~M.} \bibnamefont{Yang}}, \bibnamefont{and} \bibinfo{author}{\bibfnamefont{Z.}~\bibnamefont{Song}}, \bibinfo{journal}{Journal of Physics: Condensed Matter} \textbf{\bibinfo{volume}{33}}, \bibinfo{pages}{365403} (\bibinfo{year}{2021}).

\bibitem[{\citenamefont{Ganguly and Maiti}(2023)}]{Ganguly2023}
\bibinfo{author}{\bibfnamefont{S.}~\bibnamefont{Ganguly}} \bibnamefont{and} \bibinfo{author}{\bibfnamefont{S.~K.} \bibnamefont{Maiti}}, \bibinfo{journal}{Scientific Reports} \textbf{\bibinfo{volume}{13}}, \bibinfo{pages}{14488} (\bibinfo{year}{2023}).

\bibitem[{\citenamefont{An et~al.}(2018)\citenamefont{An, Meier, and Gadway}}]{An2018}
\bibinfo{author}{\bibfnamefont{F.~A.} \bibnamefont{An}}, \bibinfo{author}{\bibfnamefont{E.~J.} \bibnamefont{Meier}}, \bibnamefont{and} \bibinfo{author}{\bibfnamefont{B.}~\bibnamefont{Gadway}}, \bibinfo{journal}{Physical Review X} \textbf{\bibinfo{volume}{8}}, \bibinfo{pages}{031045} (\bibinfo{year}{2018}).

\bibitem[{\citenamefont{Nandkishore}(2015)}]{Nandkishore2015}
\bibinfo{author}{\bibfnamefont{R.}~\bibnamefont{Nandkishore}}, \bibinfo{journal}{Physical Review B} \textbf{\bibinfo{volume}{92}}, \bibinfo{pages}{245141} (\bibinfo{year}{2015}).

\bibitem[{\citenamefont{Hyatt et~al.}(2017)\citenamefont{Hyatt, Garrison, Potter, and Bauer}}]{Hyatt2017}
\bibinfo{author}{\bibfnamefont{K.}~\bibnamefont{Hyatt}}, \bibinfo{author}{\bibfnamefont{J.~R.} \bibnamefont{Garrison}}, \bibinfo{author}{\bibfnamefont{A.~C.} \bibnamefont{Potter}}, \bibnamefont{and} \bibinfo{author}{\bibfnamefont{B.}~\bibnamefont{Bauer}}, \bibinfo{journal}{Physical Review B} \textbf{\bibinfo{volume}{95}}, \bibinfo{pages}{035132} (\bibinfo{year}{2017}).

\bibitem[{\citenamefont{Rubio-Abadal et~al.}(2019)\citenamefont{Rubio-Abadal, Choi, Zeiher, Hollerith, Rui, Bloch, and Gross}}]{RubioAbadal2019}
\bibinfo{author}{\bibfnamefont{A.}~\bibnamefont{Rubio-Abadal}}, \bibinfo{author}{\bibfnamefont{J.-y.} \bibnamefont{Choi}}, \bibinfo{author}{\bibfnamefont{J.}~\bibnamefont{Zeiher}}, \bibinfo{author}{\bibfnamefont{S.}~\bibnamefont{Hollerith}}, \bibinfo{author}{\bibfnamefont{J.}~\bibnamefont{Rui}}, \bibinfo{author}{\bibfnamefont{I.}~\bibnamefont{Bloch}}, \bibnamefont{and} \bibinfo{author}{\bibfnamefont{C.}~\bibnamefont{Gross}}, \bibinfo{journal}{Physical Review X} \textbf{\bibinfo{volume}{9}}, \bibinfo{pages}{041014} (\bibinfo{year}{2019}).

\bibitem[{\citenamefont{Zhang et~al.}(2010)\citenamefont{Zhang, Yang, Zhao, Duan, Zhang, and Ulloa}}]{Zhang2010}
\bibinfo{author}{\bibfnamefont{W.}~\bibnamefont{Zhang}}, \bibinfo{author}{\bibfnamefont{R.}~\bibnamefont{Yang}}, \bibinfo{author}{\bibfnamefont{Y.}~\bibnamefont{Zhao}}, \bibinfo{author}{\bibfnamefont{S.}~\bibnamefont{Duan}}, \bibinfo{author}{\bibfnamefont{P.}~\bibnamefont{Zhang}}, \bibnamefont{and} \bibinfo{author}{\bibfnamefont{S.~E.} \bibnamefont{Ulloa}}, \bibinfo{journal}{Physical Review B} \textbf{\bibinfo{volume}{81}}, \bibinfo{pages}{214202} (\bibinfo{year}{2010}).

\bibitem[{\citenamefont{Lin and Gong}(2024)}]{Lin2024}
\bibinfo{author}{\bibfnamefont{X.}~\bibnamefont{Lin}} \bibnamefont{and} \bibinfo{author}{\bibfnamefont{M.}~\bibnamefont{Gong}}, \bibinfo{journal}{Physical Review A} \textbf{\bibinfo{volume}{109}}, \bibinfo{pages}{033310} (\bibinfo{year}{2024}).

\bibitem[{\citenamefont{Goswami et~al.}(2025)\citenamefont{Goswami, Chatterjee, Modak, and Sahoo}}]{Goswami2025}
\bibinfo{author}{\bibfnamefont{A.}~\bibnamefont{Goswami}}, \bibinfo{author}{\bibfnamefont{P.}~\bibnamefont{Chatterjee}}, \bibinfo{author}{\bibfnamefont{R.}~\bibnamefont{Modak}}, \bibnamefont{and} \bibinfo{author}{\bibfnamefont{S.}~\bibnamefont{Sahoo}}, \bibinfo{journal}{Physical Review B} \textbf{\bibinfo{volume}{112}}, \bibinfo{pages}{144205} (\bibinfo{year}{2025}).

\bibitem[{\citenamefont{Roati et~al.}(2008)\citenamefont{Roati, D'Errico, Fallani, Fattori, Fort, Zaccanti, Modugno, Modugno, and Inguscio}}]{Roati2008}
\bibinfo{author}{\bibfnamefont{G.}~\bibnamefont{Roati}}, \bibinfo{author}{\bibfnamefont{C.}~\bibnamefont{D'Errico}}, \bibinfo{author}{\bibfnamefont{L.}~\bibnamefont{Fallani}}, \bibinfo{author}{\bibfnamefont{M.}~\bibnamefont{Fattori}}, \bibinfo{author}{\bibfnamefont{C.}~\bibnamefont{Fort}}, \bibinfo{author}{\bibfnamefont{M.}~\bibnamefont{Zaccanti}}, \bibinfo{author}{\bibfnamefont{G.}~\bibnamefont{Modugno}}, \bibinfo{author}{\bibfnamefont{M.}~\bibnamefont{Modugno}}, \bibnamefont{and} \bibinfo{author}{\bibfnamefont{M.}~\bibnamefont{Inguscio}}, \bibinfo{journal}{Nature} \textbf{\bibinfo{volume}{453}}, \bibinfo{pages}{895} (\bibinfo{year}{2008}).

\bibitem[{\citenamefont{Billy et~al.}(2008)\citenamefont{Billy, Josse, Zuo, Bernard, Hambrecht, Lugan, Cl{\'e}ment, Sanchez-Palencia, Bouyer, and Aspect}}]{Billy2008}
\bibinfo{author}{\bibfnamefont{J.}~\bibnamefont{Billy}}, \bibinfo{author}{\bibfnamefont{V.}~\bibnamefont{Josse}}, \bibinfo{author}{\bibfnamefont{Z.}~\bibnamefont{Zuo}}, \bibinfo{author}{\bibfnamefont{A.}~\bibnamefont{Bernard}}, \bibinfo{author}{\bibfnamefont{B.}~\bibnamefont{Hambrecht}}, \bibinfo{author}{\bibfnamefont{P.}~\bibnamefont{Lugan}}, \bibinfo{author}{\bibfnamefont{D.}~\bibnamefont{Cl{\'e}ment}}, \bibinfo{author}{\bibfnamefont{L.}~\bibnamefont{Sanchez-Palencia}}, \bibinfo{author}{\bibfnamefont{P.}~\bibnamefont{Bouyer}}, \bibnamefont{and} \bibinfo{author}{\bibfnamefont{A.}~\bibnamefont{Aspect}}, \bibinfo{journal}{Nature} \textbf{\bibinfo{volume}{453}}, \bibinfo{pages}{891} (\bibinfo{year}{2008}).

\bibitem[{\citenamefont{Lahini et~al.}(2009)\citenamefont{Lahini, Pugatch, Pozzi, Sorel, Morandotti, Davidson, and Silberberg}}]{Lahini2009}
\bibinfo{author}{\bibfnamefont{Y.}~\bibnamefont{Lahini}}, \bibinfo{author}{\bibfnamefont{R.}~\bibnamefont{Pugatch}}, \bibinfo{author}{\bibfnamefont{F.}~\bibnamefont{Pozzi}}, \bibinfo{author}{\bibfnamefont{M.}~\bibnamefont{Sorel}}, \bibinfo{author}{\bibfnamefont{R.}~\bibnamefont{Morandotti}}, \bibinfo{author}{\bibfnamefont{N.}~\bibnamefont{Davidson}}, \bibnamefont{and} \bibinfo{author}{\bibfnamefont{Y.}~\bibnamefont{Silberberg}}, \bibinfo{journal}{Physical Review Letters} \textbf{\bibinfo{volume}{103}}, \bibinfo{pages}{013901} (\bibinfo{year}{2009}).

\bibitem[{\citenamefont{Schwartz et~al.}(2007)\citenamefont{Schwartz, Bartal, Fishman, and Segev}}]{Schwartz2007}
\bibinfo{author}{\bibfnamefont{T.}~\bibnamefont{Schwartz}}, \bibinfo{author}{\bibfnamefont{G.}~\bibnamefont{Bartal}}, \bibinfo{author}{\bibfnamefont{S.}~\bibnamefont{Fishman}}, \bibnamefont{and} \bibinfo{author}{\bibfnamefont{M.}~\bibnamefont{Segev}}, \bibinfo{journal}{Nature} \textbf{\bibinfo{volume}{446}}, \bibinfo{pages}{52} (\bibinfo{year}{2007}).

\bibitem[{\citenamefont{Dom{\'{}}nguez-Castro and Paredes}(2019)}]{DominguezCastro2019}
\bibinfo{author}{\bibfnamefont{G.~A.} \bibnamefont{Dom{\'{}}nguez-Castro}} \bibnamefont{and} \bibinfo{author}{\bibfnamefont{R.}~\bibnamefont{Paredes}}, \bibinfo{journal}{European Journal of Physics} \textbf{\bibinfo{volume}{40}}, \bibinfo{pages}{045403} (\bibinfo{year}{2019}).

\bibitem[{\citenamefont{Rossignolo and Dell'Anna}(2019{\natexlab{b}})}]{PhysRevB.99.054211}
\bibinfo{author}{\bibfnamefont{M.}~\bibnamefont{Rossignolo}} \bibnamefont{and} \bibinfo{author}{\bibfnamefont{L.}~\bibnamefont{Dell'Anna}}, \bibinfo{journal}{Phys. Rev. B} \textbf{\bibinfo{volume}{99}}, \bibinfo{pages}{054211} (\bibinfo{year}{2019}{\natexlab{b}}).

\bibitem[{def()}]{definition}
\bibinfo{note}{For the leg-resolved analysis, we define $\mathrm{IPR}_{m,\ell}=\sum_{i=1}^{L}|\psi_m(\ell,i)|^4$ for leg-$\ell$ ($\ell=1,2$). The corresponding eigenstate-averaged quantity is $\langle\mathrm{IPR}\rangle_\ell= \frac{1}{N}\sum_{m=1}^{N}\mathrm{IPR}_{m,\ell}$, with $N=2L$.}

\bibitem[{\citenamefont{Roy et~al.}(2018)\citenamefont{Roy, Khaymovich, Das, and Moessner}}]{Roy2018multifractality}
\bibinfo{author}{\bibfnamefont{S.}~\bibnamefont{Roy}}, \bibinfo{author}{\bibfnamefont{I.~M.} \bibnamefont{Khaymovich}}, \bibinfo{author}{\bibfnamefont{A.}~\bibnamefont{Das}}, \bibnamefont{and} \bibinfo{author}{\bibfnamefont{R.}~\bibnamefont{Moessner}}, \bibinfo{journal}{SciPost Physics} \textbf{\bibinfo{volume}{4}}, \bibinfo{pages}{025} (\bibinfo{year}{2018}).

\bibitem[{\citenamefont{Yao et~al.}(2019)\citenamefont{Yao, Khoudli, Bresque, and Sanchez-Palencia}}]{PhysRevLett.123.070405}
\bibinfo{author}{\bibfnamefont{H.}~\bibnamefont{Yao}}, \bibinfo{author}{\bibfnamefont{A.}~\bibnamefont{Khoudli}}, \bibinfo{author}{\bibfnamefont{L.}~\bibnamefont{Bresque}}, \bibnamefont{and} \bibinfo{author}{\bibfnamefont{L.}~\bibnamefont{Sanchez-Palencia}}, \bibinfo{journal}{Phys. Rev. Lett.} \textbf{\bibinfo{volume}{123}}, \bibinfo{pages}{070405} (\bibinfo{year}{2019}).

\bibitem[{\citenamefont{Hiramoto and Abe}(1988)}]{hiramoto1988dynamics2}
\bibinfo{author}{\bibfnamefont{H.}~\bibnamefont{Hiramoto}} \bibnamefont{and} \bibinfo{author}{\bibfnamefont{S.}~\bibnamefont{Abe}}, \bibinfo{journal}{Journal of the Physical Society of Japan} \textbf{\bibinfo{volume}{57}}, \bibinfo{pages}{1365} (\bibinfo{year}{1988}).

\bibitem[{\citenamefont{Giri et~al.}(2026)\citenamefont{Giri, Tai-Ming, Sengupta, and Chen}}]{SuppMat}
\bibinfo{author}{\bibfnamefont{M.~K.} \bibnamefont{Giri}}, \bibinfo{author}{\bibfnamefont{S.}~\bibnamefont{Tai-Ming}}, \bibinfo{author}{\bibfnamefont{P.}~\bibnamefont{Sengupta}}, \bibnamefont{and} \bibinfo{author}{\bibfnamefont{P.}~\bibnamefont{Chen}} (\bibinfo{year}{2026}), \bibinfo{note}{see Supplemental Material for additional numerical calculations, long-time dynamics, and circuit transpilation details}.

\bibitem[{\citenamefont{Luitz et~al.}(2014)\citenamefont{Luitz, Alet, and Laflorencie}}]{PhysRevLett.112.057203}
\bibinfo{author}{\bibfnamefont{D.~J.} \bibnamefont{Luitz}}, \bibinfo{author}{\bibfnamefont{F.}~\bibnamefont{Alet}}, \bibnamefont{and} \bibinfo{author}{\bibfnamefont{N.}~\bibnamefont{Laflorencie}}, \bibinfo{journal}{Phys. Rev. Lett.} \textbf{\bibinfo{volume}{112}}, \bibinfo{pages}{057203} (\bibinfo{year}{2014}).

\bibitem[{\citenamefont{Li et~al.}(2023)\citenamefont{Li, Wang, Shi, Huang, Song, Liang, Mei, Zhou, Zhang, Zhang et~al.}}]{li2023observation}
\bibinfo{author}{\bibfnamefont{H.}~\bibnamefont{Li}}, \bibinfo{author}{\bibfnamefont{Y.-Y.} \bibnamefont{Wang}}, \bibinfo{author}{\bibfnamefont{Y.-H.} \bibnamefont{Shi}}, \bibinfo{author}{\bibfnamefont{K.}~\bibnamefont{Huang}}, \bibinfo{author}{\bibfnamefont{X.}~\bibnamefont{Song}}, \bibinfo{author}{\bibfnamefont{G.-H.~s.} \bibnamefont{Liang}}, \bibinfo{author}{\bibfnamefont{Z.-Y.} \bibnamefont{Mei}}, \bibinfo{author}{\bibfnamefont{B.}~\bibnamefont{Zhou}}, \bibinfo{author}{\bibfnamefont{H.}~\bibnamefont{Zhang}}, \bibinfo{author}{\bibfnamefont{J.-C.} \bibnamefont{Zhang}}, \bibnamefont{et~al.}, \bibinfo{journal}{npj Quantum Information} \textbf{\bibinfo{volume}{9}}, \bibinfo{pages}{40} (\bibinfo{year}{2023}).

\bibitem[{\citenamefont{Giri and Mandal}(2025)}]{Giri_2025}
\bibinfo{author}{\bibfnamefont{M.~K.} \bibnamefont{Giri}} \bibnamefont{and} \bibinfo{author}{\bibfnamefont{S.~B.} \bibnamefont{Mandal}}, \bibinfo{journal}{Physica Scripta} \textbf{\bibinfo{volume}{100}}, \bibinfo{pages}{035111} (\bibinfo{year}{2025}), ISSN \bibinfo{issn}{1402-4896}.

\bibitem[{\citenamefont{Luo and Zhang}(2004)}]{hellinger_1}
\bibinfo{author}{\bibfnamefont{S.}~\bibnamefont{Luo}} \bibnamefont{and} \bibinfo{author}{\bibfnamefont{Q.}~\bibnamefont{Zhang}}, \bibinfo{journal}{Physical Review A} \textbf{\bibinfo{volume}{69}}, \bibinfo{pages}{032106} (\bibinfo{year}{2004}).

\bibitem[{\citenamefont{Di~Bartolomeo et~al.}(2023)\citenamefont{Di~Bartolomeo, Vischi, Cesa, Wixinger, Grossi, Donadi, and Bassi}}]{hellinger_2}
\bibinfo{author}{\bibfnamefont{G.}~\bibnamefont{Di~Bartolomeo}}, \bibinfo{author}{\bibfnamefont{M.}~\bibnamefont{Vischi}}, \bibinfo{author}{\bibfnamefont{F.}~\bibnamefont{Cesa}}, \bibinfo{author}{\bibfnamefont{R.}~\bibnamefont{Wixinger}}, \bibinfo{author}{\bibfnamefont{M.}~\bibnamefont{Grossi}}, \bibinfo{author}{\bibfnamefont{S.}~\bibnamefont{Donadi}}, \bibnamefont{and} \bibinfo{author}{\bibfnamefont{A.}~\bibnamefont{Bassi}}, \bibinfo{journal}{Phys. Rev. Res.} \textbf{\bibinfo{volume}{5}}, \bibinfo{pages}{043210} (\bibinfo{year}{2023}).

\bibitem[{\citenamefont{Lloyd}(1996)}]{Lloyd1996}
\bibinfo{author}{\bibfnamefont{S.}~\bibnamefont{Lloyd}}, \bibinfo{journal}{Science} \textbf{\bibinfo{volume}{273}}, \bibinfo{pages}{1073} (\bibinfo{year}{1996}).

\bibitem[{\citenamefont{Childs et~al.}(2018)\citenamefont{Childs, Maslov, Nam, Ross, and Su}}]{Childs2018}
\bibinfo{author}{\bibfnamefont{A.~M.} \bibnamefont{Childs}}, \bibinfo{author}{\bibfnamefont{D.}~\bibnamefont{Maslov}}, \bibinfo{author}{\bibfnamefont{Y.}~\bibnamefont{Nam}}, \bibinfo{author}{\bibfnamefont{N.~J.} \bibnamefont{Ross}}, \bibnamefont{and} \bibinfo{author}{\bibfnamefont{Y.}~\bibnamefont{Su}}, \bibinfo{journal}{Proceedings of the National Academy of Sciences} \textbf{\bibinfo{volume}{115}}, \bibinfo{pages}{9456} (\bibinfo{year}{2018}).

\bibitem[{\citenamefont{Childs et~al.}(2021)\citenamefont{Childs, Su, Tran, Wiebe, and Zhu}}]{Childs2021}
\bibinfo{author}{\bibfnamefont{A.~M.} \bibnamefont{Childs}}, \bibinfo{author}{\bibfnamefont{Y.}~\bibnamefont{Su}}, \bibinfo{author}{\bibfnamefont{M.~C.} \bibnamefont{Tran}}, \bibinfo{author}{\bibfnamefont{N.}~\bibnamefont{Wiebe}}, \bibnamefont{and} \bibinfo{author}{\bibfnamefont{S.}~\bibnamefont{Zhu}}, \bibinfo{journal}{Physical Review X} \textbf{\bibinfo{volume}{11}}, \bibinfo{pages}{011020} (\bibinfo{year}{2021}).

\bibitem[{\citenamefont{van~der Wiel et~al.}(2002)\citenamefont{van~der Wiel, De~Franceschi, Elzerman, Fujisawa, Tarucha, and Kouwenhoven}}]{vanDerWiel2002}
\bibinfo{author}{\bibfnamefont{W.~G.} \bibnamefont{van~der Wiel}}, \bibinfo{author}{\bibfnamefont{S.}~\bibnamefont{De~Franceschi}}, \bibinfo{author}{\bibfnamefont{J.~M.} \bibnamefont{Elzerman}}, \bibinfo{author}{\bibfnamefont{T.}~\bibnamefont{Fujisawa}}, \bibinfo{author}{\bibfnamefont{S.}~\bibnamefont{Tarucha}}, \bibnamefont{and} \bibinfo{author}{\bibfnamefont{L.~P.} \bibnamefont{Kouwenhoven}}, \bibinfo{journal}{Reviews of Modern Physics} \textbf{\bibinfo{volume}{75}}, \bibinfo{pages}{1} (\bibinfo{year}{2002}).

\bibitem[{\citenamefont{Burkard et~al.}(2023)\citenamefont{Burkard, Ladd, Pan, Nichol, and Petta}}]{Burkard2023}
\bibinfo{author}{\bibfnamefont{G.}~\bibnamefont{Burkard}}, \bibinfo{author}{\bibfnamefont{T.~D.} \bibnamefont{Ladd}}, \bibinfo{author}{\bibfnamefont{A.}~\bibnamefont{Pan}}, \bibinfo{author}{\bibfnamefont{J.~M.} \bibnamefont{Nichol}}, \bibnamefont{and} \bibinfo{author}{\bibfnamefont{J.~R.} \bibnamefont{Petta}}, \bibinfo{journal}{Reviews of Modern Physics} \textbf{\bibinfo{volume}{95}}, \bibinfo{pages}{025003} (\bibinfo{year}{2023}).

\end{thebibliography}


\begin{thebibliography}{13}
\expandafter\ifx\csname natexlab\endcsname\relax\def\natexlab#1{#1}\fi
\expandafter\ifx\csname bibnamefont\endcsname\relax
  \def\bibnamefont#1{#1}\fi
\expandafter\ifx\csname bibfnamefont\endcsname\relax
  \def\bibfnamefont#1{#1}\fi
\expandafter\ifx\csname citenamefont\endcsname\relax
  \def\citenamefont#1{#1}\fi
\expandafter\ifx\csname url\endcsname\relax
  \def\url#1{\texttt{#1}}\fi
\expandafter\ifx\csname urlprefix\endcsname\relax\def\urlprefix{URL }\fi
\providecommand{\bibinfo}[2]{#2}
\providecommand{\eprint}[2][]{\url{#2}}

\bibitem[{\citenamefont{Ortiz et~al.}(2001)\citenamefont{Ortiz, Gubernatis,
  Knill, and Laflamme}}]{S-Ortiz2001}
\bibinfo{author}{\bibfnamefont{G.}~\bibnamefont{Ortiz}},
  \bibinfo{author}{\bibfnamefont{J.~E.} \bibnamefont{Gubernatis}},
  \bibinfo{author}{\bibfnamefont{E.}~\bibnamefont{Knill}}, \bibnamefont{and}
  \bibinfo{author}{\bibfnamefont{R.}~\bibnamefont{Laflamme}},
  \bibinfo{journal}{Physical Review A} \textbf{\bibinfo{volume}{64}},
  \bibinfo{pages}{022319} (\bibinfo{year}{2001}).

\bibitem[{\citenamefont{Jiang et~al.}(2018)\citenamefont{Jiang, Sung,
  Kechedzhi, Smelyanskiy, and Boixo}}]{S-Jiang2018}
\bibinfo{author}{\bibfnamefont{Z.}~\bibnamefont{Jiang}},
  \bibinfo{author}{\bibfnamefont{K.~J.} \bibnamefont{Sung}},
  \bibinfo{author}{\bibfnamefont{K.}~\bibnamefont{Kechedzhi}},
  \bibinfo{author}{\bibfnamefont{V.~N.} \bibnamefont{Smelyanskiy}},
  \bibnamefont{and} \bibinfo{author}{\bibfnamefont{S.}~\bibnamefont{Boixo}},
  \bibinfo{journal}{Physical Review Applied} \textbf{\bibinfo{volume}{9}},
  \bibinfo{pages}{044036} (\bibinfo{year}{2018}).

\bibitem[{\citenamefont{Hatano and Suzuki}(2005)}]{S-suzuki_trotter}
\bibinfo{author}{\bibfnamefont{N.}~\bibnamefont{Hatano}} \bibnamefont{and}
  \bibinfo{author}{\bibfnamefont{M.}~\bibnamefont{Suzuki}}, in
  \emph{\bibinfo{booktitle}{{Quantum annealing and other optimization
  methods}}} (\bibinfo{publisher}{Springer}, \bibinfo{year}{2005}), pp.
  \bibinfo{pages}{37--68}.

\bibitem[{\citenamefont{Childs et~al.}(2021)\citenamefont{Childs, Su, Tran,
  Wiebe, and Zhu}}]{S-Childs2021}
\bibinfo{author}{\bibfnamefont{A.~M.} \bibnamefont{Childs}},
  \bibinfo{author}{\bibfnamefont{Y.}~\bibnamefont{Su}},
  \bibinfo{author}{\bibfnamefont{M.~C.} \bibnamefont{Tran}},
  \bibinfo{author}{\bibfnamefont{N.}~\bibnamefont{Wiebe}}, \bibnamefont{and}
  \bibinfo{author}{\bibfnamefont{S.}~\bibnamefont{Zhu}},
  \bibinfo{journal}{Physical Review X} \textbf{\bibinfo{volume}{11}},
  \bibinfo{pages}{011020} (\bibinfo{year}{2021}).

\bibitem[{\citenamefont{Smith et~al.}(2019)\citenamefont{Smith, Kim, Pollmann,
  and Knolle}}]{S-Smith2019}
\bibinfo{author}{\bibfnamefont{A.}~\bibnamefont{Smith}},
  \bibinfo{author}{\bibfnamefont{M.~S.} \bibnamefont{Kim}},
  \bibinfo{author}{\bibfnamefont{F.}~\bibnamefont{Pollmann}}, \bibnamefont{and}
  \bibinfo{author}{\bibfnamefont{J.}~\bibnamefont{Knolle}},
  \bibinfo{journal}{npj Quantum Information} \textbf{\bibinfo{volume}{5}},
  \bibinfo{pages}{106} (\bibinfo{year}{2019}), ISSN \bibinfo{issn}{2056-6387}.

\bibitem[{\citenamefont{Giri and Mandal}(2025)}]{S-Giri_2025}
\bibinfo{author}{\bibfnamefont{M.~K.} \bibnamefont{Giri}} \bibnamefont{and}
  \bibinfo{author}{\bibfnamefont{S.~B.} \bibnamefont{Mandal}},
  \bibinfo{journal}{Physica Scripta} \textbf{\bibinfo{volume}{100}},
  \bibinfo{pages}{035111} (\bibinfo{year}{2025}), ISSN
  \bibinfo{issn}{1402-4896}.

\bibitem[{\citenamefont{Kanti~Giri and Chen}(2026)}]{S-giri_qst}
\bibinfo{author}{\bibfnamefont{M.}~\bibnamefont{Kanti~Giri}} \bibnamefont{and}
  \bibinfo{author}{\bibfnamefont{P.}~\bibnamefont{Chen}},
  \bibinfo{journal}{Quantum Science and Technology}
  \textbf{\bibinfo{volume}{11}}, \bibinfo{pages}{035064}
  (\bibinfo{year}{2026}).

\bibitem[{\citenamefont{Gray}(2018)}]{S-Gray2018}
\bibinfo{author}{\bibfnamefont{J.}~\bibnamefont{Gray}},
  \bibinfo{journal}{Journal of Open Source Software}
  \textbf{\bibinfo{volume}{3}}, \bibinfo{pages}{819} (\bibinfo{year}{2018}).

\bibitem[{\citenamefont{Byrd et~al.}(1995)\citenamefont{Byrd, Lu, Nocedal, and
  Zhu}}]{S-Byrd1995}
\bibinfo{author}{\bibfnamefont{R.~H.} \bibnamefont{Byrd}},
  \bibinfo{author}{\bibfnamefont{P.}~\bibnamefont{Lu}},
  \bibinfo{author}{\bibfnamefont{J.}~\bibnamefont{Nocedal}}, \bibnamefont{and}
  \bibinfo{author}{\bibfnamefont{C.}~\bibnamefont{Zhu}}, \bibinfo{journal}{SIAM
  Journal on Scientific Computing} \textbf{\bibinfo{volume}{16}},
  \bibinfo{pages}{1190} (\bibinfo{year}{1995}).

\bibitem[{\citenamefont{Aleksandrowicz et~al.}(2019)}]{S-qiskit}
\bibinfo{author}{\bibfnamefont{G.}~\bibnamefont{Aleksandrowicz}}
  \bibnamefont{et~al.}, \emph{\bibinfo{title}{Qiskit: An open-source framework
  for quantum computing}} (\bibinfo{year}{2019}),
  \urlprefix\url{https://doi.org/10.5281/zenodo.2562111}.

\bibitem[{\citenamefont{Viola et~al.}(1999)\citenamefont{Viola, Knill, and
  Lloyd}}]{S-Viola1999}
\bibinfo{author}{\bibfnamefont{L.}~\bibnamefont{Viola}},
  \bibinfo{author}{\bibfnamefont{E.}~\bibnamefont{Knill}}, \bibnamefont{and}
  \bibinfo{author}{\bibfnamefont{S.}~\bibnamefont{Lloyd}},
  \bibinfo{journal}{Physical Review Letters} \textbf{\bibinfo{volume}{82}},
  \bibinfo{pages}{2417} (\bibinfo{year}{1999}).

\bibitem[{\citenamefont{Wallman and Emerson}(2016)}]{S-Wallman2016}
\bibinfo{author}{\bibfnamefont{J.~J.} \bibnamefont{Wallman}} \bibnamefont{and}
  \bibinfo{author}{\bibfnamefont{J.}~\bibnamefont{Emerson}},
  \bibinfo{journal}{Physical Review A} \textbf{\bibinfo{volume}{94}},
  \bibinfo{pages}{052325} (\bibinfo{year}{2016}).

\bibitem[{\citenamefont{van~den Berg et~al.}(2022)\citenamefont{van~den Berg,
  Minev, and Temme}}]{S-vandenBerg2022}
\bibinfo{author}{\bibfnamefont{E.}~\bibnamefont{van~den Berg}},
  \bibinfo{author}{\bibfnamefont{Z.~K.} \bibnamefont{Minev}}, \bibnamefont{and}
  \bibinfo{author}{\bibfnamefont{K.}~\bibnamefont{Temme}},
  \bibinfo{journal}{Physical Review A} \textbf{\bibinfo{volume}{105}},
  \bibinfo{pages}{032620} (\bibinfo{year}{2022}).

\end{thebibliography}
%%%%%%%%%%%%%%%%%%%%%%%%%%%%%%%%%%%%%%%%%%%%%%%%%%%%

\section*{\Large End Matter}
\label{sec:endmatter}
\renewcommand{\thesubsection}{EM\arabic{subsection}}
\makeatletter\renewcommand{\p@subsection}{}\makeatother % no section prefix in \ref -> "EM1", not " EM1"
\setcounter{subsection}{0}

\subsection{Real-Space Signatures}
\label{sec:E1_Di}
\begin{figure}[h!]
    \centering
    %\vspace{0.15cm}
    \includegraphics[width =1\columnwidth]{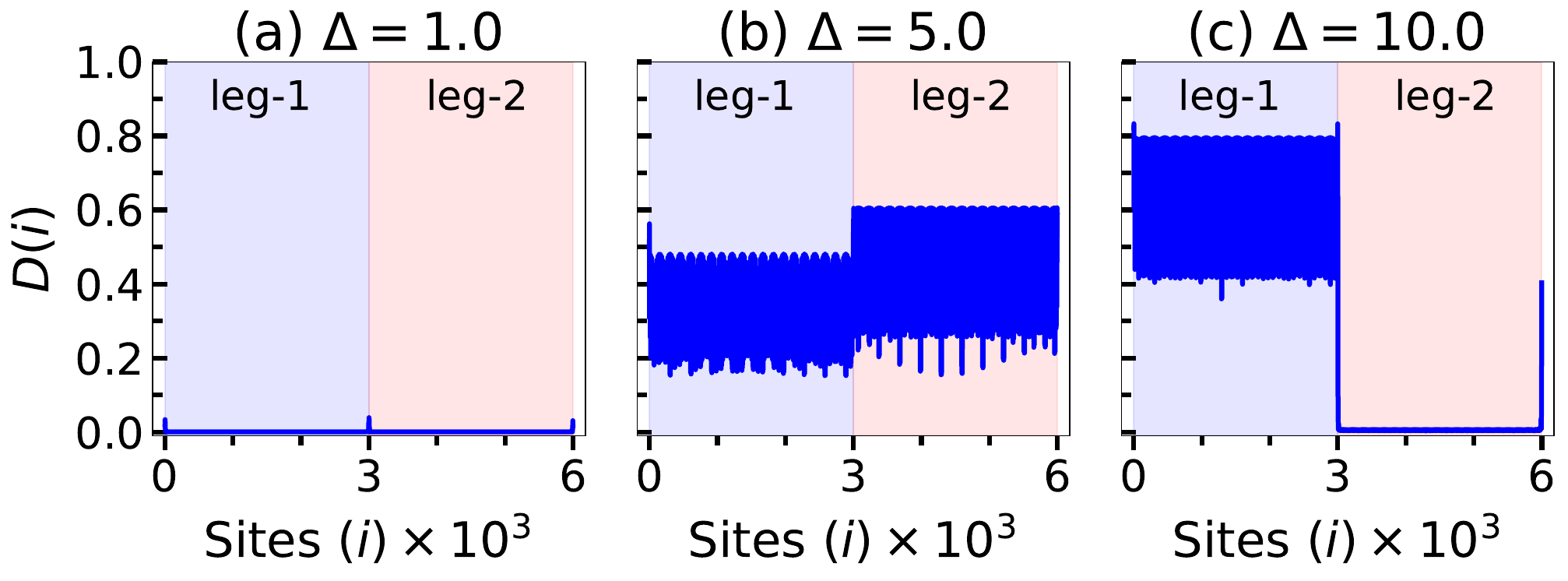}
    \caption{(a-c)The Figure shows the site-resolved $D(i)$ for $\Delta=1.0$, $5.0$ and $10.0$, respectively. Here we consider length of each chain, $L=3000$, $\lambda=0.5$ and $K=3.0$. }
    %\vspace{-0.5cm}
    \label{fig:psi_plot}
\end{figure}

To provide real-space evidence for proximity-induced localization (PIL) and the subsequent re-entrant regime, we examine the site-resolved accumulated weight
$D(i)=\sum_n |\psi_n(i)|^4$
across the two legs of the ladder. 
This quantity provides a spectrum-wide real-space diagnostic of localization: extended states yield nearly vanishing $D(i)$, whereas localized states exhibit enhanced $D(i)$ due to the spatial concentration of eigenstate amplitudes.
%This quantity acts as a spectrum-wide real-space diagnostic: small values distributed over the bulk indicate extended eigenstates, whereas a large plateau reflects the accumulation of localized eigenstates over a finite spatial region.
Figure~\ref{fig:psi_plot} shows $D(i)$ for representative detunings $\Delta=1.0$, $5.0$, and $10.0$ at $K=3.0$ and $\lambda=0.5$.
%For $\Delta=1.0$, $D(i)$ remains small across both legs, consistent with predominantly extended states. At $\Delta=5.0$, $D(i)$ increases strongly on both legs and forms a broad oscillatory plateau, demonstrating simultaneous localization of the two legs.
At $\Delta=1.0$, $D(i)$ remains nearly vanishing across both legs, consistent with predominantly extended states. At $\Delta=5.0$, both legs exhibit enhanced $D(i)$, indicating simultaneous localization and providing real-space evidence of PIL on leg-2, which is consistent with the energy spectrum in Fig.~\ref{fig:phase_diagram}(e) and the leg-resolved $\langle\mathrm{IPR}\rangle_\ell$ behavior in Fig.~\ref{fig:IPR_leg_a_b}(b).
%At larger detuning, $\Delta=10.0$, the system enters the second IP. Leg-1 retains a broad high-amplitude profile, indicating strong localization. In contrast, leg-2 shows strongly suppressed bulk values of $D(i)$, together with a pronounced boundary peak. This behavior is consistent with the re-entrant recovery of delocalized bulk character on leg-2, accompanied by boundary-localized contributions.
At larger detuning, $\Delta=10.0$, leg-1 retains enhanced $D(i)$, indicating persistent localization. In contrast, leg-2 exhibits nearly vanishing bulk values of $D(i)$ together with a pronounced boundary peak, consistent with the coexistence of extended-like bulk states and boundary-localized states in the re-entrant regime (see \ref{sec:mixed_loc}).\\

\subsection{Projected-state analysis}
\label{sec:E2_projected}
To examine the robustness of the leg-resolved localization behavior, we perform a projected-state analysis by selecting eigenstates with dominant support on each leg. For the $n$-th eigenstate, the probability weight on leg-$\ell$ is defined as
$$
W_{n,\ell}=\sum_{i=1}^{L}|\psi_{n,\ell}(i)|^2,
$$
where $\psi_{n,\ell}(i)$ denotes the eigenstate amplitude at site $i$ of leg-$\ell$. Following Ref.~\cite{Goswami2025}, we introduce a threshold $\epsilon=0.5$ and retain only eigenstates satisfying $W_{n,\ell}>\epsilon$, ensuring that the selected states carry majority probability weight on the corresponding leg. The projected $\langle\mathrm{IPR}\rangle_\ell$ and $\langle\mathrm{NPR}\rangle_\ell$ are then evaluated by averaging over the selected eigenstates.
\begin{figure}[t]
    \centering
    %\vspace{0.15cm}
    \includegraphics[width =1\columnwidth]{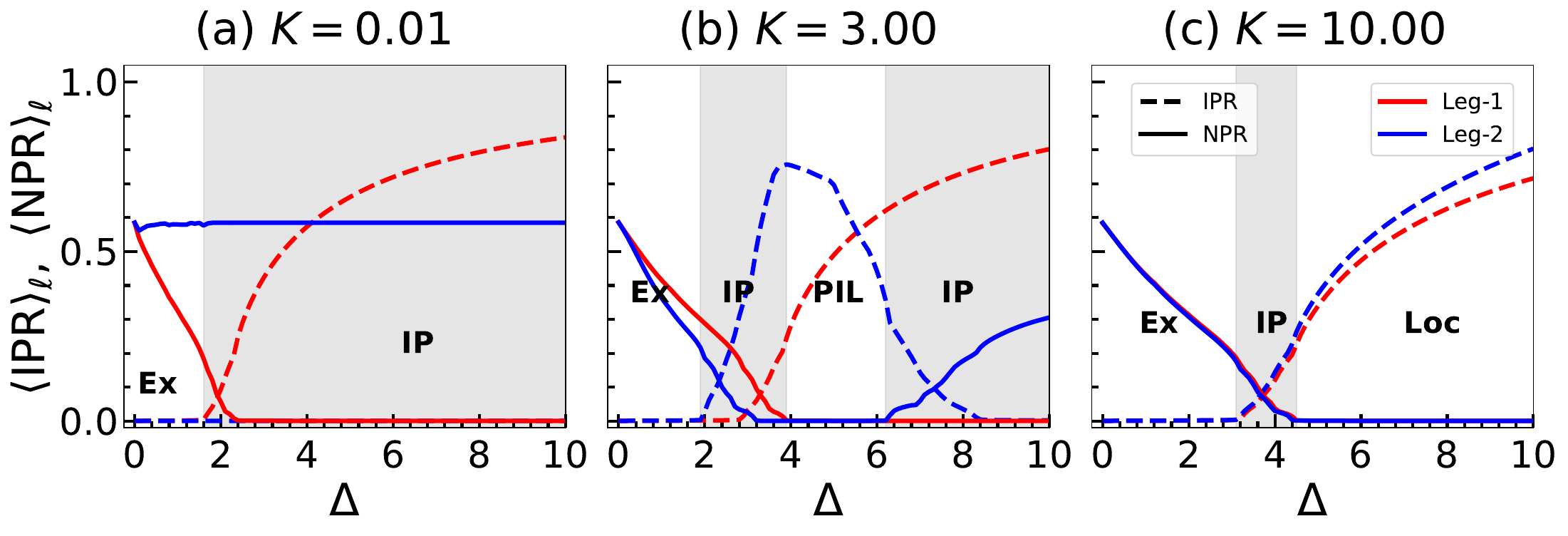}
    \caption{ Projected-state leg-resolved $\langle \mathrm{IPR}\rangle_{\ell}$ and $\langle \mathrm{NPR}\rangle_{\ell}$ as a function of $\Delta$ for threshold $\epsilon=0.5$ at $\lambda=0.5$ and $L=3000$. Panels (a)-(c) correspond to $K=0.01$, $3.0$, and $10.0$, respectively. The projected analysis reproduces the same weak-coupling IP, intermediate-coupling PIL and re-entrant IP, and strong-coupling localization sequence obtained from the full leg-resolved averages shown in main text.}
    %\vspace{-0.5cm}
    \label{fig:proj_ipr}
\end{figure}
The resulting projected $\langle\mathrm{IPR}\rangle_\ell$ and $\langle\mathrm{NPR}\rangle_\ell$ are shown in Fig.~\ref{fig:proj_ipr}. The projected analysis reproduces the weak-coupling IP, intermediate-coupling PIL regime, and large-detuning re-entrant IP observed in main text Fig.~\ref{fig:IPR_leg_a_b}, confirming that the localization behavior is robust against the choice of leg-resolved averaging procedure.

%For the projected-state analysis, we first compute the total probability weight of each eigenstate on leg-1 and leg-2. We then introduce a threshold $\epsilon$ to select states with dominant support on a given leg: an eigenstate contributes to the projected average of leg-$\ell$ only if its weight on that leg exceeds $\epsilon$. Here we use $\epsilon=0.5$, so that the retained states carry majority weight on the corresponding leg. This projection filters out weakly supported leg components and provides a stricter measure of leg-resolved localization~\cite{Goswami2025}.
%The projected $\langle \mathrm{IPR}\rangle_{\ell}$ and $\langle \mathrm{NPR}\rangle_{\ell}$ are shown in Fig.~\ref{fig:proj_ipr}. The resulting phase structure is consistent with the full leg-resolved analysis, confirming that the weak-coupling IP, the intermediate-coupling PIL window, and the large-detuning re-entrant IP are robust against the choice of leg-resolved averaging procedure.

\subsection{Mixed localization regime}
\label{sec:mixed_loc}
\begin{figure}[t]
    \centering
    %\vspace{0.15cm}
    \includegraphics[width =1.0\columnwidth]{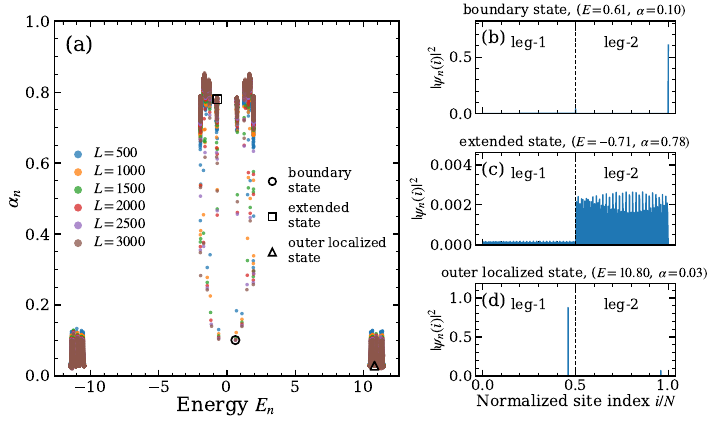}
    \caption{ State-resolved analysis at large detuning. (a) Finite-size effective scaling exponent, $\alpha_n$, with $N$, plotted versus eigen energy $E_n$ for $K=3.00$, $\lambda=0.50$, and $\Delta=10.00$. Panels (b)-(d) show the probability distribution $|\psi_n(i)|^2$ of the marked states. The dashed line separates the two legs.}
    %\vspace{-0.5cm}
    \label{fig:tau_study}
\end{figure}

%To clarify the origin of the intermediate averaged exponent $\bar{\tau}_{\ell}$ in the large-detuning regime, we perform a state-resolved analysis at $\Delta=10$. For each eigenstate, we define the finite-size effective exponent or fractal dimension~\cite{PhysRevLett.123.070405}
To clarify the origin of the intermediate averaged exponent $\bar{\tau}_\ell$ in the large-detuning regime, we perform a state-resolved analysis at $\Delta=10$. For each eigenstate, we define the finite-size effective scaling exponent~\cite{PhysRevLett.123.070405},
$$
\alpha_n=-\frac{\log_{10}(\mathrm{IPR}_n)}{\log_{10}N}, \qquad N=2L,
$$
where $\mathrm{IPR}_n=\sum_i |\psi_n(i)|^4$. Here $\alpha_n$ is used as a state-resolved indicator of the spatial character of individual eigenstates: larger $\alpha_n$ corresponds to more extended-like states, whereas smaller $\alpha_n$ corresponds to localized states. 
Figure~\ref{fig:tau_study}(a) shows that, at $\Delta=10$, the state-resolved exponent $\alpha_n$ is strongly energy dependent. Distinct spectral sectors show different behavior. The outer spectral branches have small $\alpha_n$, and representative wave-function profiles show strong localization, mainly on the detuned leg-1 sector (see Fig.~\ref{fig:tau_study}(d)). In contrast, the central spectral region contains states with relatively large $\alpha_n$ whose probability density is spread over leg-2, indicating an extended-like character (see Fig.~\ref{fig:tau_study}(c)). At the same time, a small number of low-$\alpha_n$ states also appear in the central region; their real-space profiles reveal sharp localization at the boundaries of leg-2 (see Fig.~\ref{fig:tau_study}(b)). Thus, the intermediate averaged exponent $\bar{\tau}_{\ell=2}\simeq0.45$ at large detuning should not be interpreted as evidence of a homogeneous critical or multifractal phase. Instead, it reflects a mixed localization regime characterized by the coexistence of extended-like bulk states and boundary-localized states on leg-2.
%Thus, the intermediate averaged value $\bar{\tau}_{\ell=2}\simeq 0.45$ at large $\Delta$ should not be interpreted as evidence for a homogeneous critical or multifractal phase. Rather, it reflects mixed phases, where extended and boundary-localized states co-exist.

\subsection{Circuit compression and hardware validation}
\label{sec:E4_ibmq_implementation}
For the $N=18$-qubit ladder, the circuit depth and two-qubit-gate count of the direct Suzuki–Trotter implementation increase with evolution time, making long-time dynamics increasingly susceptible to hardware errors. To address this limitation, we employ tensor-network-assisted variational circuit compression to approximate the Trotter-evolved target state at each evolution time using a bounded-depth parameterized quantum circuit. The compression accuracy and hardware performance are assessed using the Hellinger fidelity~\cite{hellinger_1,hellinger_2},
\begin{equation}
\mathcal{F}^{(\alpha)}(t) =
\left[ \sum_{x} \sqrt{ p_{x}^{\mathrm{sim}}(t) p_{x}^{(\alpha)}(t) } \right]^{2},
\label{eq:hellinger_fidelity}
\end{equation}
where $p_{x}^{\mathrm{sim}}(t)$ denotes the reference simulator probability distribution, $p_{x}^{(\alpha)}(t)$ is the distribution obtained using implementation $\alpha$. Here, $\alpha$ represents the noiseless simulation of the compressed quantum circuits (CQC), or the IBMQ execution of either the uncompressed quantum circuits (UQCs) or the CQCs. 
The Hellinger fidelity satisfies $0\leq \mathcal{F}\leq1$, with unity corresponding to identical probability distributions.
\begin{figure}[t]
    \centering
    \includegraphics[width=1.0\columnwidth]{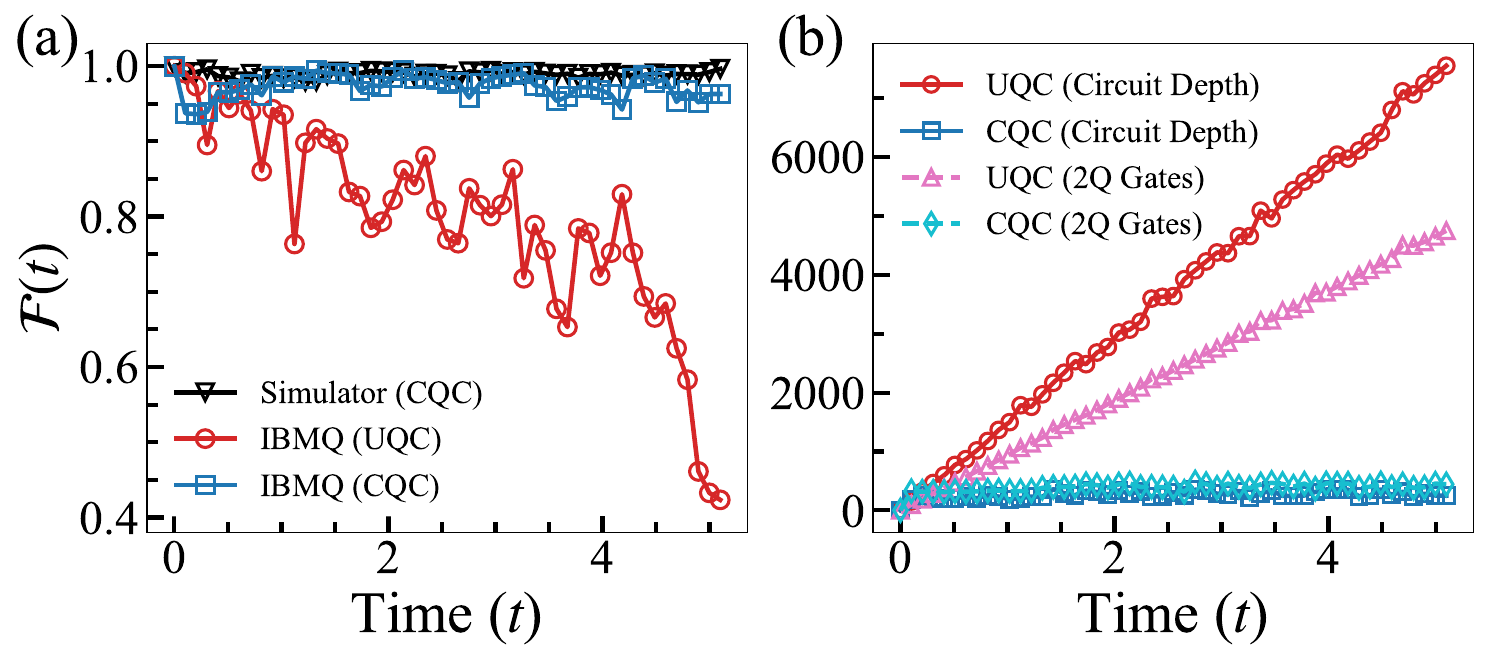}
    \caption{Validation of the tensor-network-assisted circuit compression for the $N=18$-qubit ladder at $J=1$, $K=3$, and $\Delta=0$. (a) Hellinger fidelity $\mathcal{F}(t)$ for the noiseless simulation of the compressed quantum circuits (CQCs) and for the IBMQ execution of the uncompressed (UQCs) and compressed circuits. (b) Transpiled circuit depth and two-qubit (2Q) gate count for the UQCs and CQCs. Circuit compression substantially reduces both resource measures and preserves higher hardware fidelity at longer evolution times.}
    \label{fig:fidelity_circuit_depth}
\end{figure}
Figure~\ref{fig:fidelity_circuit_depth}(a) presents a representative comparison for $J=1$, $K=3$, and $\Delta=0$. The simulated CQCs maintain $\mathcal{F}\gtrsim0.98$ throughout the reported time interval, showing that the compressed circuit accurately reproduces the reference probability distribution. On the quantum processor, the CQCs retain an average fidelity $\bar{\mathcal{F}}\approx0.97$ over this interval (and $\bar{\mathcal{F}}\approx0.95$ averaged over all hardware runs with $0\leq\Delta\leq10$), whereas the fidelity of the UQCs decreases progressively with evolution time and reaches approximately $0.42$ at $t=5J^{-1}$. The higher fidelity of the CQCs demonstrates that circuit compression reduces accumulated hardware errors, enabling more reliable quantum simulation over the accessible evolution time.
%The substantially higher fidelity of the CQC demonstrates that circuit compression suppresses the accumulation of hardware errors and extends the evolution time accessible on the quantum processor.
\begin{figure}[t]
    \includegraphics[width=0.9\columnwidth]{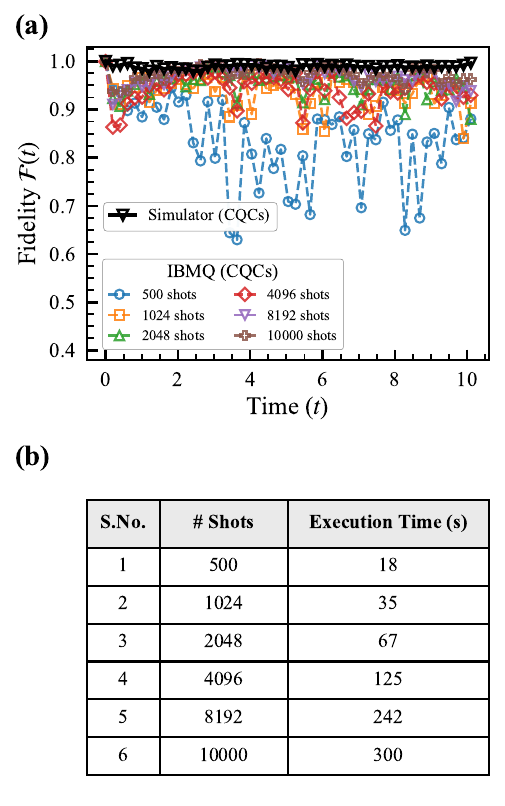}
    \caption{Shot-count dependence of the compressed-circuit execution. (a) Time-dependent Hellinger fidelity, $\mathcal{F}(t)$, obtained from the ideal simulator and IBMQ executions with shot counts ranging from $500$ to $10000$. (b) Corresponding QPU execution time for each shot count.}
    \label{fig:shots_table}
\end{figure}
The corresponding circuit resources are shown in
Fig.~\ref{fig:fidelity_circuit_depth}(b). For the UQC, both the
transpiled circuit depth and the number of two-qubit gates increase approximately linearly with time, reaching about $7.6\times10^{3}$ and $4.8\times10^{3}$, respectively, at $t=5J^{-1}$. In contrast, the CQCs remain below approximately $4\times10^{2}$ in depth and $5\times10^{2}$ two-qubit gates over the same time interval. This corresponds to reductions of approximately $95\%$ in circuit depth and $90\%$ in the two-qubit-gate count at the longest time considered. The UQCs and CQCs are transpiled using identical backend, optimization level, and transpiler settings. Complete details of the circuit construction, compression optimization, transpilation, hardware calibration, post-selection, and fidelity evaluation are provided in the SM~\cite{SuppMat}.
%%%%%%%%%%%%%%%%%%%%%%%%%%%%%%%

\subsection{Shot-count convergence and QPU runtime}
\label{sec:E6_shots_study}
%To assess the trade-off between statistical accuracy and hardware cost, we executed representative compressed circuits using $500$, $1024$, $2048$, $4096$, $8192$, and $10000$ shots. 
To assess the trade-off between measurement precision and QPU execution time, we examine the shot-count dependence of the Hellinger fidelity using representative compressed circuits with $500$, $1024$, $2048$, $4096$, $8192$, and $10000$ shots.
As shown in Fig.~\ref{fig:shots_table}, increasing the shot count progressively suppresses sampling fluctuations in the Hellinger fidelity defined in Eq.~\ref{eq:hellinger_fidelity}, with the results approaching convergence at $8192$ - $10000$ shots. The QPU execution time, however, increases from approximately $18$s for $500$ shots to $300$s for $10000$ shots. We therefore use $10000$ shots for all IBMQ results reported in this work to ensure statistically stable measurements while maintaining a feasible QPU runtime.

%%%%%%%%%%%%%%%%%%%%%%%%%%%%%%
\subsection{Trotter and hardware-error analysis}
\label{sec:E5_error_analysis}

%To identify the origin of the deviation in the Leg-2 participation entropy at $\Delta=5$, we calculate the leg-resolved density error as
To investigate the deviation in the leg-2 participation entropy at $\Delta=5$ and distinguish Trotterization errors from hardware-induced deviations, we examine the leg-resolved density error, defined as 
\begin{equation}
\epsilon_{\ell}(t)=\frac{1}{2}\sum_{i=1}^{L}\left|n_{\ell,\mathrm{Trotter}}(i,t)-n_{\ell,\mathrm{Exact}}(i,t)\right|,\qquad \ell=1,2.
\end{equation}
For $\delta t=0.10$, $\epsilon_2$ reaches approximately $0.185$ at $t=10J^{-1}$, whereas $\epsilon_1$ remains below $0.09$. 
The error decreases systematically as $\delta t$ is reduced to $0.05$ and $0.02$, indicating that finite-step Trotterization is the dominant source of the observed deviation in the ideal simulation.
%The error decreases systematically when $\delta t$ is reduced to $0.05$ and $0.02$, confirming that the deviation primarily originates from finite-step Trotterization. 
%For a first-order trotterization, the leading error introduced at each Trotter step is of order $\mathcal{O}(\delta t^2)$, while the error measured at time $t$ contains the cumulative contribution of all Trotter steps composing the evolution up to that time~\cite{Childs2021}.
For first-order Trotterization, the leading local operator error introduced at each Trotter step is of order $\mathcal{O}(\delta t^2)$. At a fixed evolution time $t$, this error accumulates over the $t/\delta t$ steps composing the evolution, resulting in a global error that is generally of order $\mathcal{O}(\delta t)$~\cite{Lloyd1996,Childs2018,Childs2021}.
%For a first-order product formula, the leading error introduced by each Trotter slice is of order $\mathcal{O}(\delta t^2)$, while the error measured at time $t$ contains the accumulated contribution of all preceding slices~\cite{Childs2021}.

%The enhanced Leg-2 error at $\Delta=5$ can be qualitatively understood as a competition between Trotter-error generation and inter-leg hybridization. 
The enhanced leg-2 error at intermediate detuning can be qualitatively understood from the competition between detuning-dependent Trotter-error generation and the suppression of inter-leg hybridization.
To identify the detuning-dependent error contribution, we write the onsite Hamiltonian
\begin{equation}
H_{\mathrm{onsite}}=\sum_x\left(V_{1,x}\hat a_x^\dagger\hat a_x+V_{2,x}\hat b_x^\dagger\hat b_x\right),
\end{equation}
where $V_{1,x}=(-1)^x\Delta+\lambda\cos(2\pi\beta x+\phi_1)$ and $V_{2,x}=\lambda\cos(2\pi\beta x+\phi_2)$. Since $H_{\mathrm{onsite}}$ does not commute with the rung-coupling term $H_{12}$, the Baker-Campbell-Hausdorff expansion gives the leading one-step correction as~\cite{Childs2021}
\begin{equation}
\Omega_{\mathrm{err}}^{(12)}=-\frac{(\delta t)^2}{2}[H_{\mathrm{onsite}},H_{12}].
\end{equation}
The commutator is controlled by the local potential difference $\delta_x=V_{1,x}-V_{2,x}$. When the staggered contribution dominates, $\delta_x\simeq(-1)^x\Delta$, and hence
\begin{equation}
\Vert\Omega_{\mathrm{err}}^{(12)}\Vert\sim K|\Delta|(\delta t)^2.
\end{equation}
This error channel is therefore weak at $\Delta=0$ and increases with $\Delta$. However, its contribution to the leg-2 density also depends on the hybridization between the two sites of each rung. Treating an individual rung as an effective two-level system gives the standard hybridization factor~\cite{vanDerWiel2002,Burkard2023} 
\begin{equation} 
    W(\Delta)=\sin^2(2\theta)=\frac{4K^2}{\Delta^2+4K^2}, 
\end{equation} 
which decreases at large detuning as the increasing spectral separation between the two leg states suppresses coherent rung hybridization. The non-monotonic leg-2 error can therefore be associated with the competition between commutator-driven error generation and the detuning-induced suppression of rung hybridization. A local-rung measure incorporating both contributions is given by
\begin{equation}
A_{\mathrm{tr}}(\Delta)\propto K|\Delta|(\delta t)^2W(\Delta).
\end{equation}
The corresponding squared magnitude of this quantity gives the following approximate detuning dependence of the leg-2 error sensitivity:
\begin{equation}
F(\Delta)\propto\left|A_{\mathrm{tr}}(\Delta)\right|^2\propto\frac{\Delta^2}{\left(\Delta^2+4K^2\right)^2}.
\end{equation}
Maximizing $F(\Delta)$ gives $\Delta_{\max}=2K$. For $K=3$, the estimated maximum therefore occurs at $\Delta_{\max}\simeq6$, close to the simulated intermediate value $\Delta=5$. This estimate is consistent with the enhanced Trotter sensitivity observed at $\Delta=5$ among the three simulated detuning values.
%Therefore, among the simulated values $\Delta=0$, $5$, and $10$, the intermediate value $\Delta=5$ is expected to exhibit the strongest Trotter sensitivity. 
At $\Delta=0$, the detuning-dependent error-generation channel is absent, whereas at $\Delta=10$, reduced rung hybridization suppresses the transmission of this error to leg-2. The resulting local-rung scaling provides a concise physical interpretation of the non-monotonic detuning dependence observed in the numerical leg-2 density error.
\begin{figure}[h!]
    %\centering
    \includegraphics[width=1.0\columnwidth]{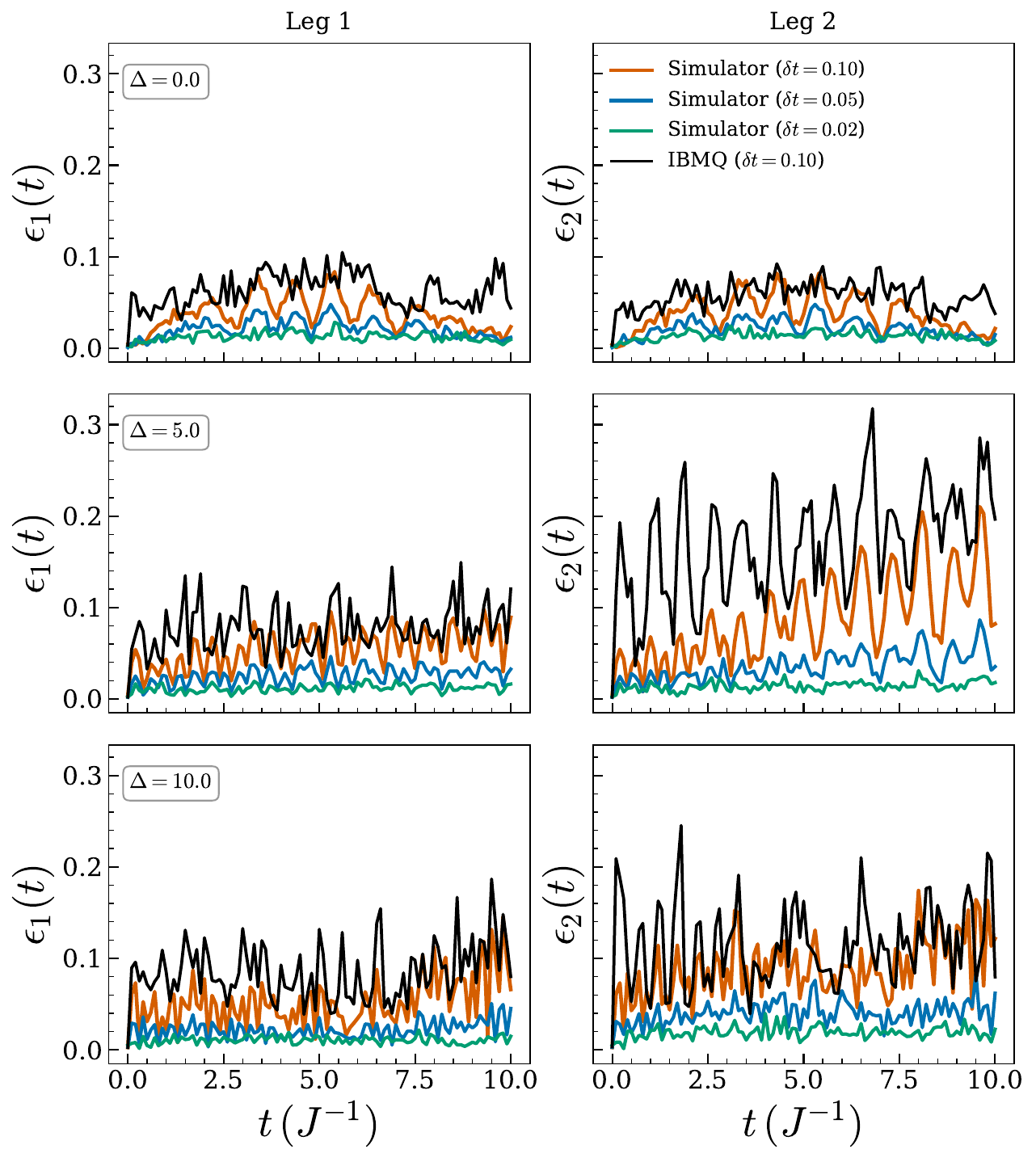}
    \caption{Leg-resolved density error $\epsilon_{\ell}(t)$ relative to the exact time evolution. The left and right columns show the errors on leg-1 and leg-2, respectively, while the rows correspond to $\Delta=0$, $5$, and $10$.}
    \label{fig:trotter_error}
\end{figure}
%\nopagebreak
%\smallskip

% ==============================================================================
% END MATTER / REFERENCES END HERE
% ==============================================================================

\pagebreak
\clearpage

% --- Supplementary Material Setup ---
\onecolumngrid % Switches to single-column layout for Supplementary Material
\begin{center}
    \textbf{\large Supplementary Material for:}\\
    \vspace{2mm}
    \textbf{\large "Quantum Simulation of Detuning-Controlled Proximity-Induced Localization and Re-entrant Delocalization in a Quasiperiodic Ladder"}
    \label{sec:supp}
\end{center}

\vspace{4mm}

% Reset equation, figure, table, and section numbering with "S" prefix
\makeatletter\setcounter{NAT@ctr}{0}\makeatother % SM references restart at S1
\renewcommand{\citenumfont}[1]{S#1}
\renewcommand{\bibnumfmt}[1]{[S#1]}
\setcounter{section}{0}
\setcounter{equation}{0}
\setcounter{figure}{0}
\setcounter{table}{0}
\setcounter{page}{1}

\renewcommand{\thesection}{S\arabic{section}}
\renewcommand{\theequation}{S\arabic{equation}}
\renewcommand{\thefigure}{S\arabic{figure}}
\renewcommand{\thetable}{S\arabic{table}}
\renewcommand{\thepage}{S\arabic{page}}
\renewcommand{\thesubsection}{S\arabic{subsection}}
\setcounter{subsection}{0}
% ==============================================================================
% SUPPLEMENTARY SECTIONS START HERE
% ==============================================================================
\section*{Details of the Implementation on IBMQ}
\label{sec:supp_exact}
\subsection{Qubit encoding and initial-state preparation}

For the digital quantum simulation, each lattice site is mapped onto a single qubit using the occupation-number encoding, where $|0\rangle$ and $|1\rangle$ represent an unoccupied and occupied site, respectively. A two-leg ladder containing $L$ sites on each leg is therefore mapped onto $N=2L$ qubits. 
%The local particle density is obtained from the Pauli-(Z) expectation value as $n_{\ell}(i,t)=\frac{1-\langle Z_{\ell,i}(t)\rangle}{2},$ where ($\ell=1,2$) labels the two legs. 
The dynamics are initialized in the single-particle sector. The particle is prepared on a selected rung $i_0$ in an equal superposition of the two legs,
\begin{equation*}
    |\Psi(0)\rangle = \frac{1}{\sqrt{2}} \left( a_{i_0}^{\dagger}+b_{i_0}^{\dagger} \right)|\mathrm{vac}\rangle.
\end{equation*} 
The two logical qubits associated with the selected rung are denoted by $q_a$ and $q_b$. Starting from $|00\rangle_{q_aq_b}$, a Hadamard gate is applied to $q_a$, followed by a controlled-NOT gate with $q_a$ as the control and $q_b$ as the target. These operations produce the Bell state ($|00\rangle+|11\rangle)/\sqrt{2}$. A subsequent Pauli-X gate on $q_b$ transforms it into the required single-particle state,

\begin{equation*}
    |\Psi(0)\rangle = 
\frac{|01\rangle_{q_aq_b}+|10\rangle_{q_aq_b}}{\sqrt{2}}.
\end{equation*}
All remaining qubits are initialized in $|0\rangle$. The corresponding preparation circuit is shown in Fig. S1(a). For the hardware simulations reported here, we use $L=9$, corresponding to a total of $N=18$ qubits.
\begin{figure}[h!]
    \centering
    \includegraphics[width=0.8\linewidth]{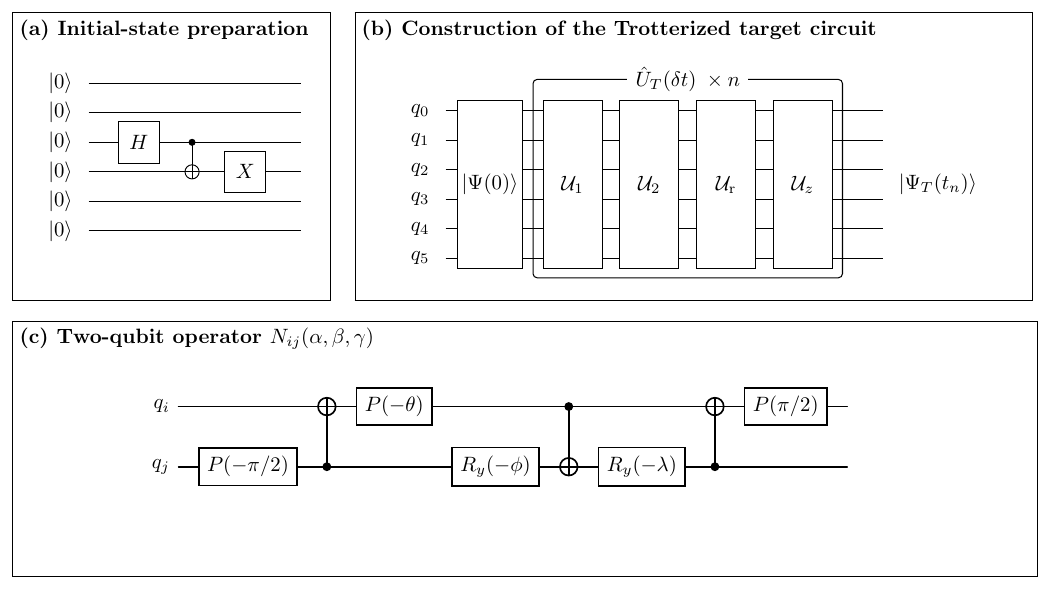}
    \caption{Circuit construction for the digital quantum simulation. (a) Preparation of the initial state. (b) Six-qubit schematic of the Trotterized target circuit. Starting from $|\Psi(0)\rangle$, each Trotter block applies the intra-leg layers $\mathcal{U}_{1}=e^{-i\hat{H}_{leg,1}}$ and $\mathcal{U}_{2}=e^{-i\hat{H}_{leg,2}}$, the inter-leg rung layer $\mathcal{U}_{\mathrm{r}}=e^{-i\hat{H}_{rung}}$, and the onsite layer $\mathcal{U}_{z}=e^{-i\hat{H}_{z}}$. Repeating this block $n$ times generates $|\Psi_T(t_n)\rangle$ at $t_n=n\delta t$. (c) Gate decomposition of the two-qubit hopping operator $\mathcal{N}_{ij}(\alpha,\beta,\gamma)$ into three CNOT gates and parameterized single-qubit rotations.}
    \label{fig:trotter_circuit}
\end{figure}
\subsection{Construction of the Trotterized target circuit}
The real-time dynamics are governed by the time-evolution operator, $U(t)=\exp(-iHt)$. To construct the corresponding quantum circuit, the ladder Hamiltonian given in main text, $H$, is first expressed in terms of Pauli operators using the occupation-number mapping~\citesupp{S-Ortiz2001,S-Jiang2018}. For two sites $i$ and $j$, the hopping operator is mapped as $\frac{1}{2} \left(X_iX_j+Y_iY_j\right),$ while the local number operator is represented by, $n_i=(1-Z_i)/2$. After omitting the constant identity contribution, which produces only a global phase, the qubit Hamiltonian takes the form
\begin{equation}
H ={} -\frac{J}{2}
\sum_{\ell=1}^{2}\sum_{i=0}^{L-2}
\left( X_{\ell,i}X_{\ell,i+1} + Y_{\ell,i}Y_{\ell,i+1} \right) 
-\frac{K}{2} \sum_{i=0}^{L-1} \left(X_{1,i}X_{2,i} + Y_{1,i}Y_{2,i} \right)
-\frac{1}{2} \sum_{\ell=1}^{2}\sum_{i=0}^{L-1} v_{\ell,i}Z_{\ell,i},
\end{equation}
Here, $J$ and $K$ denote the intra-leg and inter-leg hopping amplitudes, respectively. The onsite potentials are given by
\begin{equation}
v_{1,i} = \lambda\cos(2\pi\beta i+\phi_{1}) + \Delta(-1)^{i},
\qquad
v_{2,i} = \lambda\cos(2\pi\beta i+\phi_{2}),
\end{equation}
where leg 1 contains the additional staggered detuning $\Delta$, while leg 2 corresponds to the unmodified Aubry--Andr\'e chain. Throughout this work, we set $\phi_1=\phi_2=0$. For the circuit implementation, the sites are ordered rung by rung. Sites are labeled $i=0,\ldots,L-1$; site $i$ on leg 1 is mapped to qubit $q_{2i}$, while site $i$ on leg 2 is mapped to qubit $q_{2i+1}$. The Hamiltonian is separated into the intra-leg hopping contributions of the two legs, the inter-leg rung coupling, and the onsite contribution:
\begin{equation}
\hat{H} = \hat{H}_{\mathrm{leg},1} + \hat{H}_{\mathrm{leg},2} + \hat{H}_{\mathrm{rung}} + \hat{H}_{z}.
\end{equation}

Since these contributions do not all commute, the evolution over one time interval $\delta t$ is approximated using a first-order Suzuki-Trotter decomposition~\citesupp{S-suzuki_trotter,S-Childs2021}. As shown in Fig.~\ref{fig:trotter_circuit}(b), the circuit layers are applied from left to right in the order
$\mathcal{U}_{1}\rightarrow\mathcal{U}_{2}\rightarrow
\mathcal{U}_{\mathrm{r}}\rightarrow\mathcal{U}_{z}$.
The corresponding Trotter step is
\begin{equation}
e^{-i\hat{H}\delta t} = \hat{U}_{T}(\delta t) + \mathcal{O}(\delta t^{2}),
\qquad
\hat{U}_{T}(\delta t) = \mathcal{U}_{z}(\delta t)\;
\mathcal{U}_{\mathrm{r}}(\delta t)\;
\mathcal{U}_{2}(\delta t)\;
\mathcal{U}_{1}(\delta t).
\end{equation}

The hopping terms are implemented using the two-qubit operator
$\mathcal{N}_{ij}(\alpha,\beta,\gamma)$~\citesupp{S-Smith2019,S-Giri_2025,S-giri_qst}. With the qubit ordering defined above and $N=2L$, the circuit layers are

\begin{equation}
\begin{aligned}
\mathcal{U}_{1}(\delta t)
&=
\prod_{i=0}^{L-2}
\mathcal{N}_{2i,\,2i+2}
\left(
-\frac{J\delta t}{2},
-\frac{J\delta t}{2},
0
\right),
\\[2mm]
\mathcal{U}_{2}(\delta t)
&=
\prod_{i=0}^{L-2}
\mathcal{N}_{2i+1,\,2i+3}
\left(
-\frac{J\delta t}{2},
-\frac{J\delta t}{2},
0
\right),
\\[2mm]
\mathcal{U}_{\mathrm{r}}(\delta t)
&=
\prod_{i=0}^{L-1}
\mathcal{N}_{2i,\,2i+1}
\left(
-\frac{K\delta t}{2},
-\frac{K\delta t}{2},
0
\right),
\\[2mm]
\mathcal{U}_{z}(\delta t)
&=
\bigotimes_{i=0}^{L-1}
\left[
R_{Z,\,2i}
\left(-v_{1,i}\delta t\right)
\otimes
R_{Z,\,2i+1}
\left(-v_{2,i}\delta t\right)
\right].
\end{aligned}
\end{equation}
Thus, $\mathcal{U}_{1}$ acts on the even-indexed qubit pairs $(0,2),(2,4),\ldots$, while $\mathcal{U}_{2}$ acts on the odd-indexed pairs $(1,3),(3,5),\ldots$. The rung layer $\mathcal{U}_{\mathrm{r}}$ acts on
$(0,1),(2,3),\ldots$. The decomposition of $\mathcal{N}_{ij}(\alpha,\beta,\gamma)$ into three CNOT gates and parameterized single-qubit phase and rotation gates is shown in Fig.~\ref{fig:trotter_circuit}(c). The rotation angles are defined as
\begin{equation}
\theta=\frac{\pi}{2}-2\gamma,
\qquad
\phi=2\alpha-\frac{\pi}{2},
\qquad
\lambda=\frac{\pi}{2}-2\beta.
\end{equation}
For the intra-leg hopping terms, $\alpha=\beta=-J\delta t/2$, whereas for the rung couplings, $\alpha=\beta=-K\delta t/2$. Since the Hamiltonian contains no density-density interaction, $\gamma=0$. The onsite contribution is implemented using $R_Z(-v_{\ell,i}\delta t)$. With the Qiskit convention
$R_Z(\theta) = \exp\left(-i\frac{\theta}{2}Z\right)$, this rotation produces the required onsite evolution $\exp(i v_{\ell,i}\delta t Z/2)$, apart from an irrelevant global phase.
Repeating the elementary Trotter block $n$ times generates the target state at
$t_n=n\delta t$:
\begin{equation}
\left|\Psi_T(t_n)\right\rangle = \left[\hat{U}_T(\delta t)\right]^n \left|\Psi(0)\right\rangle.
\end{equation}
We use $\delta t=0.1J^{-1}$ and evolve the system up to $t=10J^{-1}$, corresponding to a maximum of 100 Trotter steps. Including the initial state at $t=0$, the dynamics are evaluated at 101 time points. A separate target state is constructed for each $t_n$ and subsequently used in the tensor-network-assisted variational compression.

\begin{figure}[t]
    \includegraphics[width=1\linewidth]{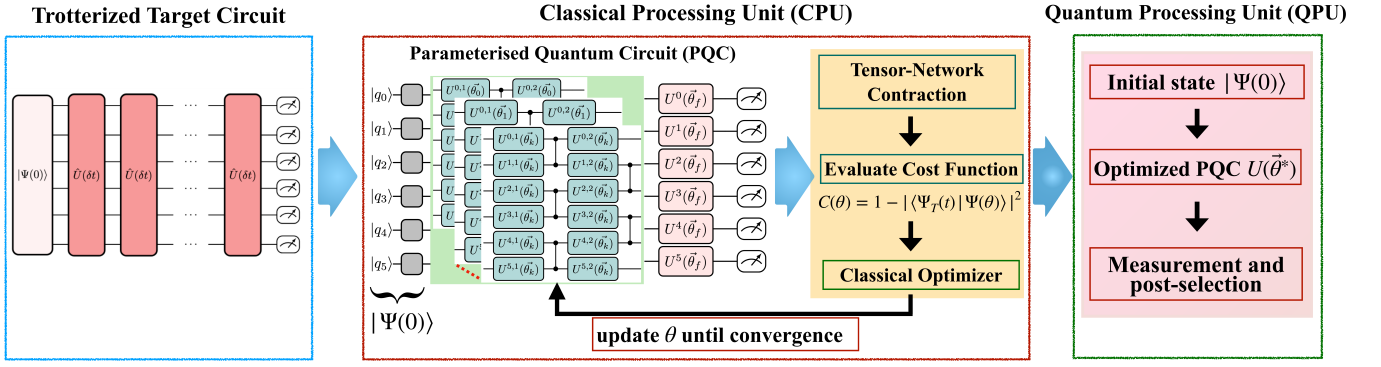}
    %\vspace{-3cm}
    \caption{Schematic of the tensor-network-assisted circuit-compression workflow. A bounded-depth parameterized quantum circuit is optimized classically to reproduce the Trotterized target state and is subsequently executed on the quantum processor.}
    \label{fig:compression}
\end{figure}
\subsection{Tensor-network-assisted circuit compression}
The depth of the Trotterized circuit increases with the number of time steps, making its direct execution increasingly sensitive to gate errors and decoherence. To obtain a shallow circuit for each evolution time $t_n$, we approximate the Trotterized target state $|\Psi_T(t_n)\rangle$ using a parameterized quantum circuit $\hat{U}_C(\boldsymbol{\theta})$. Acting on the same initial state, the compressed circuit generates
\begin{equation}
|\Psi_C(t_n;\boldsymbol{\theta})\rangle
=
\hat{U}_C(\boldsymbol{\theta})|\Psi(0)\rangle .
\end{equation}
The ansatz consists of parameterized single-qubit unitaries followed by alternating layers of nearest-neighbour two-qubit unitaries arranged in a brick-wall structure, as illustrated in Fig.~\ref{fig:compression}. A final layer of single-qubit unitaries is included to increase the flexibility of the ansatz. Unlike the Trotterized circuit, whose depth increases with $t_n$, the variational circuit has a bounded depth determined by the accuracy required to represent the target state.

The Trotterized target circuit and the parameterized circuit are represented as tensor networks on the classical processing unit. Their contraction directly evaluates the overlap between the target and variational states. The circuit parameters are optimized by minimizing the cost function
\begin{equation}
C(\boldsymbol{\theta};t_n)
=
1-
\left|
\langle\Psi_T(t_n)|
\Psi_C(t_n;\boldsymbol{\theta})\rangle
\right|^2 .
\end{equation}
The corresponding compression fidelity is defined as
\begin{equation}
F_{\mathrm{c}}(t_n)
=
\left|
\langle\Psi_T(t_n)|
\Psi_C(t_n;\boldsymbol{\theta}^{*})\rangle
\right|^2 ,
\end{equation}
where $\boldsymbol{\theta}^{*}$ denotes the optimized parameters. The tensor-network contractions and parameter optimization are performed using the QUIMB library~\citesupp{S-Gray2018}. The gradients of the cost function are obtained through automatic differentiation, and the parameters are updated using the L-BFGS-B optimizer~\citesupp{S-Byrd1995}. If the optimization saturates before reaching the desired fidelity, the ansatz depth is increased by one layer and the optimization is repeated. A separate optimization is carried out for every evolution time $t_n$. Therefore, the compressed circuit is state-specific: it approximates the Trotter-evolved state produced from the chosen initial state rather than the complete time-evolution operator for an arbitrary input state. The optimization is performed entirely on the classical processor, and no quantum-hardware data are used to determine the circuit parameters. After convergence, the optimized circuit $\hat{U}_C(\boldsymbol{\theta}^{*})$ is transpiled and executed on the quantum processor~\citesupp{S-giri_qst}.

The compression accuracy is quantified by the state-overlap fidelity $F_{\mathrm{c}}(t)$ between the compressed and Trotterized target states. The compressed circuits are further validated at the probability-distribution level using the Hellinger fidelity. As detailed in \ref{sec:E4_ibmq_implementation}, noiseless and hardware executions of the uncompressed quantum circuits (UQCs) and compressed quantum circuits (CQCs) are compared under identical transpilation and hardware settings. At $t=5J^{-1}$, the UQCs reach a transpiled depth of approximately $7.6\times10^{3}$ with $4.8\times10^{3}$ two-qubit gates, whereas the CQCs remain below approximately $4\times10^{2}$ in depth and $5\times10^{2}$ two-qubit gates. This corresponds to reductions of approximately $95\%$ in circuit depth and $90\%$ in the two-qubit-gate count.

\subsection{Transpilation, Error Suppression and hardware execution}

The circuits were transpiled and executed on the IBMQ backend \texttt{ibm\_boston}, a 156-qubit Heron r3 processor. Transpilation was performed using Qiskit with optimization level 3, including hardware-aware qubit-layout selection, routing, gate cancellation, and decomposition into the native gate set~\citesupp{S-qiskit}. A connected subset of $18$ physical qubits was selected to reduce routing overhead and accumulated two-qubit-gate errors. The backend calibration reported a layered two-qubit-gate error of $\sim2.64\times10^{-3}$, with median and best two-qubit-gate errors of $\sim1.43\times10^{-3}$ and $\sim6.18\times10^{-4}$, respectively. The same backend and transpilation settings were used for the uncompressed and compressed circuits to ensure a consistent comparison. Each circuit was executed using $10000$ measurement shots. All qubits were measured in the computational basis. 

Hardware errors were suppressed using dynamical decoupling and Pauli twirling implemented through the Qiskit Runtime Sampler. The target time-evolution operator $U(t)$ is approximated by the compressed-circuit evolution $U_{\mathrm{C}}(t)$. Dynamical-decoupling pulses were inserted only during idle intervals of the scheduled circuit, thereby suppressing idle-qubit dephasing without modifying $U_{\mathrm{C}}(t)$~\citesupp{S-Viola1999}. Denoting the unwanted evolution during an idle interval by $U_{\mathrm{noise}}(\tau)$, an $X$ pulse reverses the phase accumulated under quasi-static $Z$-type dephasing, such that $XU_{\mathrm{noise}}(\tau)X=U_{\mathrm{noise}}^{\dagger}(\tau)$. The ideal $XX$ refocusing block therefore satisfies 
\begin{equation} 
    D_{XX}(2\tau)=XU_{\mathrm{noise}}(\tau)XU_{\mathrm{noise}}(\tau)=U_{\mathrm{noise}}^{\dagger}(\tau)U_{\mathrm{noise}}(\tau)=I. 
\end{equation} 
Thus, the unwanted phases accumulated during successive idle intervals cancel to leading order, while the intended compressed-circuit evolution remains unchanged.

Pauli twirling was applied to two-qubit gates to reduce the systematic accumulation of coherent gate errors~\citesupp{S-Wallman2016}. The physical noise channel $\mathcal{E}$ is averaged over randomly selected operators from the Pauli group $\mathcal{P}$ according to \begin{equation} \mathcal{E}_{\mathrm{twirl}}(\rho) = \frac{1}{|\mathcal{P}|} \sum_{P\in\mathcal{P}} P^{\dagger}\mathcal{E}\!\left(P\rho P^{\dagger}\right)P, \end{equation} which transforms the averaged noise into an effectively stochastic Pauli channel without changing the ideal circuit operation. Measurement twirling was additionally applied to the terminal measurements to symmetrize the readout channel~\citesupp{S-vandenBerg2022}. The resulting measurement distributions were subsequently processed using the post-selection described in the following section.

\subsection{Measurement and post-selection}

The initial state and the Hamiltonian conserve the total particle number, which remains equal to one throughout the ideal evolution. Hardware errors can, however, produce measurement outcomes outside this physical sector. We therefore apply post-selection by retaining only the measured bitstrings with Hamming weight one. The physical single-particle sector is defined as
\begin{equation}
\mathcal{S}_{1}
=
\left\{
x\,\middle|\, w(x)=\sum_{j=1}^{N}x_j=1
\right\}.
\end{equation}
If $N_x(t)$ is the number of occurrences of the bitstring $x$ at time $t$, the total number of retained shots is
\begin{equation}
N_{\mathrm{ps}}(t)
=
\sum_{x\in\mathcal{S}_{1}}N_x(t).
\end{equation}
The probabilities of the accepted outcomes are then renormalized as
\begin{equation}
p_x^{\mathrm{ps}}(t)
=
\frac{N_x(t)}{N_{\mathrm{ps}}(t)},
\qquad x\in\mathcal{S}_{1}.
\end{equation}
%The retained-shot fraction is calculated as $f_{\mathrm{keep}}(t)=N_{\mathrm{ps}}(t)/N_{\mathrm{shots}}$, where $N_{\mathrm{shots}}=4096$. 
All local densities and dynamical observables are evaluated using the post-selected probability distribution. The same post-selection procedure is applied to the uncompressed and compressed hardware results.

%\section{Long-time evolution results}
%\label{sec:supp_trotter}
%\bibliographysupp{rho}            % Supplementary 
\bibliographystylesupp{apsrev}
\bibliographysupp{PR_supp}
\end{document}